\documentclass{ieeeaccess}
\usepackage{cite}
\usepackage{amsmath,amssymb,amsfonts}
\usepackage{algorithmic}
\usepackage{graphicx}
\usepackage{textcomp}
\def\BibTeX{{\rm B\kern-.05em{\sc i\kern-.025em b}\kern-.08em
    T\kern-.1667em\lower.7ex\hbox{E}\kern-.125emX}}
    
\usepackage[table]{xcolor}
\usepackage{caption}
\usepackage{braket}
\usepackage{bm}
\usepackage{tikz}
\usetikzlibrary{quantikz}
\NewSpotColorSpace{PANTONE}
  \AddSpotColor{PANTONE} {PANTONE3015C} {PANTONE\SpotSpace 3015\SpotSpace C} {1 0.3 0 0.2}
  \SetPageColorSpace{PANTONE}%
\usepackage{algorithm}
\usepackage{multirow}
\usepackage[colorlinks=true,citecolor=blue,linkcolor=magenta]{hyperref}

\def\textsc#1{\textnormal{{\sc #1}}}
\newcommand{\Gate}[1]{\textsc{#1}}

\newcommand{\zgate}{\Gate{z}}
\newcommand{\ygate}{\Gate{y}}
\newcommand{\xgate}{\Gate{x}}

\newcommand{\tgate}{\Gate{t}}

\newcommand{\idgate}{\Gate{i}}

\newcommand{\cnotgate}{\Gate{cnot}}

\newcommand{\rygate}{\Gate{r}_y}
\newcommand{\rzgate}{\Gate{r}_z}

\begin{document}
\history{Date of publication xxxx 00, 0000, date of current version xxxx 00, 0000.}
\doi{10.1109/TQE.2020.DOI}

\title{Layer VQE: A Variational Approach for Combinatorial Optimization on Noisy~Quantum Computers}

\author{\uppercase{Xiaoyuan Liu}\authorrefmark{1}, \IEEEmembership{Member, IEEE},
\uppercase{Anthony Angone}\authorrefmark{2},
\uppercase{Ruslan Shaydulin}\authorrefmark{3}, \IEEEmembership{Member, IEEE}, 
\uppercase{Ilya Safro}\authorrefmark{1},
\uppercase{Yuri Alexeev}\authorrefmark{4}, 
\IEEEmembership{Senior Member, IEEE},
and \uppercase{Lukasz Cincio}.\authorrefmark{5}}

\address[1]{Department of Computer and Information Sciences, University of Delaware, Newark, DE 19716, USA}
\address[2]{School of Computing, Clemson University, Clemson, SC 29634, USA}
\address[3]{Mathematics and Computer Science Division, Argonne National Laboratory, Lemont, IL 60439, USA}
\address[4]{Computational Science Division, Argonne National Laboratory, Lemont, IL 60439, USA}
\address[5]{Theoretical Division, Los Alamos National Laboratory, Los Alamos, NM 87545, USA}
\tfootnote{X.L., A.A., R.S., I.S. and Y.A. were supported in part with funding from the Defense Advanced Research Projects Agency (DARPA). R.S. and Y.A. were supported by Laboratory Directed Research and Development (LDRD) funding from Argonne National Laboratory, provided by the Director, Office of Science, of the U.S. Department of Energy under Contract No. DE-AC02-06CH11357. R.S. was supported by the U.S. Department of Energy, Office of Science, Office of Advanced Scientific Computing Research, Accelerated Research for Quantum Computing program. L.C. was supported by the Laboratory Directed Research and Development (LDRD) program of Los Alamos National Laboratory (LANL) under project number 20200056DR. LANL is operated by Triad National Security, LLC, for the National Nuclear Security Administration of U.S. Department of Energy (contract no. 89233218CNA000001). L.C. was also supported by the U.S. DOE, Office of Science, Office of Advanced Scientific Computing Research, under the Accelerated Research in Quantum Computing (ARQC) program.}

\markboth
{Author \headeretal: Preparation of Papers for IEEE Transactions on Quantum Engineering}
{Author \headeretal: Preparation of Papers for IEEE Transactions on Quantum Engineering}

\corresp{Corresponding author: Xiaoyuan Liu (email: joeyxliu@udel.edu).}

\begin{abstract}
Combinatorial optimization on near-term quantum devices is a promising path to demonstrating quantum advantage. However, the capabilities of these devices are constrained by high noise or error rates. In this paper, we propose an iterative Layer VQE (L-VQE) approach, inspired by the Variational Quantum Eigensolver (VQE). We present a large-scale numerical study, simulating circuits with up to 40 qubits and 352 parameters, that demonstrates the potential of the proposed approach. We evaluate quantum optimization heuristics on the problem of detecting multiple communities in networks, for which we introduce a novel qubit-frugal formulation. We numerically compare L-VQE with Quantum Approximate Optimization Algorithm (QAOA) and demonstrate that QAOA achieves lower approximation ratios while requiring significantly deeper circuits. We show that L-VQE is more robust to finite sampling errors and has a higher chance of finding the solution as compared with standard VQE approaches. Our simulation results show that L-VQE performs well under realistic hardware noise.
\end{abstract}


\titlepgskip=-15pt

\maketitle

\section{Introduction}

Recent advances in quantum computing hardware open the possibility of demonstrating quantum advantage in practical applications~\cite{alexeev2019quantum,shaydulin2019hybrid}. A promising target application domain is combinatorial optimization, with problems becoming classically intractable (in the current state of theory) to solve exactly even for moderately sized instances. This situation suggests that the requirement for the number of qubits needed to tackle certain classically hard combinatorial optimization problems is relatively low, leading to the possibility of noisy intermediate-scale quantum (NISQ)~\cite{preskill2018quantum} devices becoming competitive with classical state-of-the-art methods for such problems.

Near-term quantum devices are expected to have high noise levels, and only partial error mitigation is currently possible. This situation leads to a constraint on the maximum depth of the quantum circuit that can be reliably executed on NISQ devices. This constraint motivated the development of a number of hybrid quantum-classical algorithms for optimization, most notably the Quantum Approximate Optimization Algorithm (QAOA)~\cite{farhi2014quantum,farhi2017quantum,hadfield2019quantum} and variational quantum algorithms for optimization~\cite{moll2018quantum,nannicini2019performance}. These algorithms execute only a short parameterized circuit on the quantum computer and use a classical outer-loop procedure to find ``good'' parameters~\cite{cerezo2020variational}. The short parameterized circuit is often referred to as the ansatz. The goal of the outer-loop procedure, in general, is to find parameters such that the output of the quantum circuit includes high-quality solutions to the combinatorial optimization problem being solved.

The choice of the ansatz is a key problem in hybrid algorithms. One major concern is the expressivity of the chosen ansatz. The ansatz has to be sufficiently expressive, meaning that there should exist parameters with which the ansatz prepares a state suitably close to the solution of the problem (note that we are not referring to the expressibility formally defined in~\cite{holmes2021connecting}). Another thing to consider is the optimization of the parameters, sufficiently good parameters have to be tractable to find~\cite{holmes2021connecting}.

For combinatorial optimization problems, the solution is classical; in other words, it is a computational basis state. Therefore, the expressivity of the ansatz, reduces to the ability to prepare a state with sufficiently large overlap with the computational basis state encoding the solution of the problem. This means that the ansatz can be sufficiently expressive without generating any entanglement or having any quantum properties whatsoever: one layer of single-qubit rotations is sufficient to prepare an arbitrary computational basis state. However, finding good parameters may be challenging for such ans\"atze. Their structure leads to localized optimization, which is prone to local minima. As we discuss below, that class of ans\"atze may be extended to enhance the ability to find good parameters by introducing a correlation between distant parts of the system. A commonly used class of highly expressive ans\"atze are those with alternating layers of single-qubit and two-qubit gates, where the two-qubit gates are aligned with the connectivity available on the hardware. These ans\"atze are known as quantum neural networks~\cite{mcclean2018barren} or hardware-efficient ans\"atze~\cite{kandala2017hardware}. An alternative and ``natural'' approach is the Hamiltonian-evolution ansatz used in QAOA.  Such ans\"atze can be less expressive, however, since the state it has to prepare is a nontrivial entangled state due to the symmetry-preserving properties of the ansatz~\cite{shaydulin2020classical}. This observation has been used by Bravyi et al.~\cite{bravyi2019obstacles} to show that because of the $\mathbb{Z}_2$ symmetry of the ansatz, QAOA with constant depth is outperformed by the classical Goemans--Williamson algorithm for MaxCut. As a result, QAOA needs a comparatively large circuit depth to achieve the same (classical) expressivity as compared with hardware-efficient ansatz.

For ans\"atze with a large number of parameters, the high-quality parameters are typically found by using a classical optimizer. Thus the second criterion, the ability to find sufficiently good parameters, is typically framed in terms of the cost function landscape that the classical outer-loop routine has to optimize over. Recent results show that highly expressive ans\"atze such as hardware-efficient ans\"atze suffer from ``barren plateaus'' in the optimization landscape, making finding high-quality parameters intractable~\cite{mcclean2018barren, cerezo20cost, sharma20trainability, pesah20absence, cerezo2021higher, wang20noise, holmes20barren, volkoff20efficient, campos2020abrupt, marrero20entanglement, abbas20power}. At the same time, a series of recent results show that because of the structured nature of the ansatz used in QAOA, one may be able to find high-quality parameters by using machine learning approaches~\cite{khairy2019learning,wilson2019optimizing,verdon2019learning} or by restricting the parameters to a specific physically motivated class~\cite{zhou2020quantum,crooks2018performance,mbeng2019quantum}.

\begin{figure*}[!htbp]
    \centering
    \includegraphics[width=\textwidth]{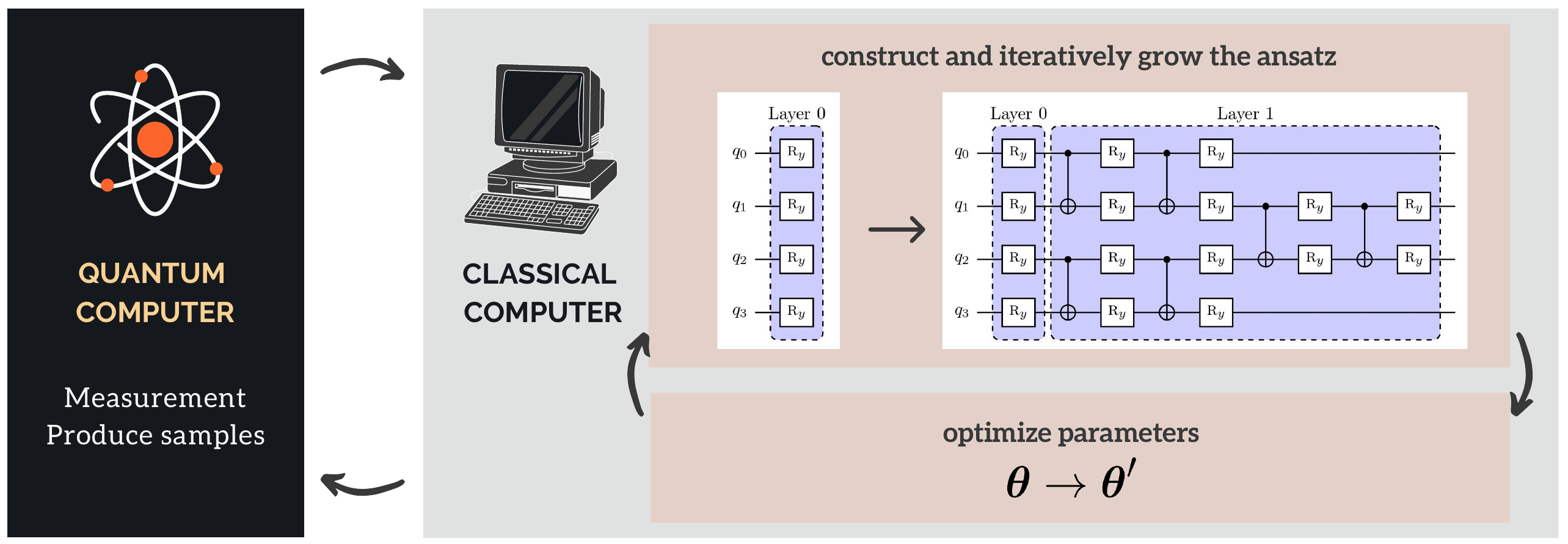}
    \caption{Layer-VQE: start from a simple and shallow ansatz with one $\rygate$ act on each qubit; optimize and update the parameters; after some predefined number of iterations; increment the size of the ansatz; optimize and update all parameters. The ansatz can be incremented multiple times.}
	\label{fig:lvqe}
\end{figure*}

In this paper, we propose a practical approach to combinatorial optimization on near-term quantum computers. We introduce an iterative approach, which we call Layer VQE (L-VQE), inspired by recent advances in hybrid quantum-classical algorithms with an adaptive ansatz~\cite{grimsley2019adaptive,zhu2020adaptive,skolik2020layerwise}. In L-VQE, we start with one layer of parameterized rotations and increment the size of the ansatz systematically by introducing entangling gates and additional parameterized rotations. To heuristically decrease the likelihood of getting trapped in a local optimum of the parameters, we increment the ansatz before reaching convergence. To guarantee that at each step the quality of the solution does not decrease, we initialize the added ansatz such that it evaluates to identity. We work with qubits aligned in a chain and assume nearest neighbor connectivity, which is a reasonable assumption as most common hardware topology include line as a subgraph. Restricting ourselves to this class of problems allows us to benchmark the proposed methods on large problems in simulation by using tensor network techniques. We expect that in practical applications on real quantum hardware, one would organize ansatz layers according to (typically two-dimensional) qubit connectivity to further enhance circuit expressiveness. Quantum circuits on such layouts cannot be in general classically efficiently simulated and are thus not considered in the present study.

Fig. \ref{fig:lvqe} gives a schematic presentation of L-VQE. We study the algorithm for the problem of detecting $k$ communities in networks, and we propose a novel qubit-frugal formulation with many-body interactions in the Hamiltonian. For a network with $n$ nodes, $n \left\lceil \log_2 k \right\rceil$ qubits are required for the circuit. We present a large-scale numerical study of the proposed approach, simulating circuits with up to 40 qubits and 352 rotational gates (i.e., parameters). Our numerical simulation results show that the proposed approach achieves a higher approximation ratio compared with QAOA while requiring significantly lower circuit depth. The proposed approach is more robust to finite sampling error (i.e., if the objective value is not known exactly and is estimated by drawing samples from the quantum state) and performs better than hybrid approaches with a fixed ansatz. Moreover, we show that the proposed approach performs well under realistic hardware noise by using a trapped ion noisy quantum simulator.

The rest of the paper is organized as follows. In Section \ref{sec:background} we review the relevant background of solving combinatorial optimizations on quantum computers. In Section \ref{sec:related} we review related work. Section \ref{sec:lvqe} introduces our L-VQE approach, and in Section \ref{sec:modularity} we discuss our novel formulation of the $k$-community detection problem. Section \ref{sec:experiments} presents our numerical simulation results and in Section~\ref{sec:conclusions} we summarize our conclusions.

\section{Background}\label{sec:background}

We begin by briefly reviewing our notion of combinatorial optimization on quantum computers and relevant concepts. Suppose we have an objective function $C(x)$ defined on the Boolean cube $x = \{x_i\}_{i=1}^n \in \{0, 1\}^n$ and a corresponding optimization problem
\begin{equation}\label{eq:optobj1}
    \max_{x \in \{0,1\}^n}{C(x)},
\end{equation}
where the objective function $C(x)$ can be formulated in the following form:

\begin{equation}\label{eq:optobjcanonical}
C(x) = \sum_{q} w_q \prod_{i\in q} x_i \prod_{j\not\in q} (1-x_j).
\end{equation}
Here, $q \subset \{1, 2, \cdots, n\}$ are given index sets, and $w_q$ are given coefficients. The objective function $C(x)$ is said to be \emph{faithfully represented} by a Hamiltonian $\mathcal{H}$ if it acts as $\mathcal{H}\ket{x}=C(x)\ket{x}$ for each $x\in\{0,1\}^n$.
For a function given in the form (\ref{eq:optobjcanonical}), such a Hamiltonian representation can be constructed by substituting every $x_i$ with the matrix $x_i \to \frac{1}{2} (\idgate - \zgate_i)$, where $\idgate$ is the identity matrix and $\zgate_i$ is the Pauli $\zgate$ operator that acts on qubit $i$:
\begin{equation}\label{eq:hamiltoniancanonical}
\mathcal{H} = \sum_{q} w_q \prod_{i\in q} \frac{\idgate - \zgate_i}{2} \prod_{j\in q^C} \frac{\idgate - \zgate_j}{2}.
\end{equation}
Note that the operator $\mathcal{H}\in\mathbb{C}^{2^n}$ is never constructed explicitly. Instead, we construct a compact representation of it as a combination of Pauli $\zgate$ operators.

\subsection{Combinatorial Optimization on Near-Term Quantum Computers}\label{sec:backgroundCOonquantum}

The two most prominent candidate algorithms for combinatorial optimization on noisy near-term quantum computers are the Variational Quantum Eigensolver (VQE, originally proposed in the context of quantum chemistry~\cite{peruzzo2014variational}) and the Quantum Approximate Optimization Algorithm (QAOA)~\cite{farhi2014quantum}. Both algorithms are hybrid quantum-classical algorithms that combine a parameterized trial state $\ket{\psi(\bm{\theta})}$ prepared on a quantum computer with a classical routine used to find high-quality parameters $\bm{\theta}$. The goal is to find parameters $\bm{\theta}$ such that when the state $\ket{\psi(\bm{\theta})}$ is measured, the measurement result corresponds to a good solution of the classical optimization problem. The parameterized trial state $\ket{\psi(\bm{\theta})}$ is commonly called the ansatz.

In VQE, for optimization the ansatz is frequently tailored to the hardware~\cite{nannicini2019performance,Quinones2020Tailored}, and the parameters $\bm{\theta}$ are found by using a classical outer-loop optimizer. The expectation value $\braket{\psi(\bm{\theta}) | \mathcal{H} | \psi(\bm{\theta})}$ is commonly used as the metric for the optimizer, although other approaches have been suggested~\cite{Barkoutsos2020}. QAOA uses a problem-dependent ansatz given by
\begin{equation}
\ket{\psi_p(\bm{\gamma}, \bm{\beta})} = e^{-i\beta_p B} e^{-i\gamma_p \mathcal{H}} \cdots e^{-i\beta_1 B} e^{-i\gamma_1 \mathcal{H}} \ket{+}^{\otimes n},
\end{equation}
where $B = \sum_{i=1}^n \xgate_i$ is the mixing Hamiltonian, $\xgate_i$ is the Pauli $\xgate$ operator acting on qubit $i$, $\mathcal{H}$ is the Hamiltonian faithfully representing the objective function, and $p$ is a parameter controlling the depth. The special structure of the QAOA ansatz enables finding high-quality parameters $\bm{\gamma}, \bm{\beta}$ purely classically in many settings~\cite{farhi2014quantum,Streif2020,2101.10296} or using very few iterations of the outer-loop optimizer~\cite{brandao2018fixed-rs,khairy2019learning,shaydulin2019multistart}.

We evaluate the quality of the final quantum state $\ket{\psi(\bm{\theta})}$ by computing the approximation ratio $\rho$ defined as follows:
\begin{equation}\label{eq:approximationratio}
\rho = \frac{\braket{\psi(\bm{\theta}) | \mathcal{H} | \psi(\bm{\theta})}}{C_{\operatorname{bkv}}},
\end{equation}
where $C_{\operatorname{bkv}}$ is either the global optimum of the objective function $C(x)$ if available, or the best known value otherwise. We defined approximation ratio with respect to the best known value since the global optimal $\max_{x \in \{0,1\}^n}{C(x)}$ may not be accessible for sufficiently large problem instances.

\subsection{The $k$-Community Detection}

The $k$-community detection, also known as modularity clustering, is a famous problem in network science. The goal is to partition network nodes into $k$ communities (also known as clusters or parts) such that the modularity metric~\cite{newman2006modularity} defined in Eq. \ref{eq:modularity} is maximized. There are several graph partitioning problems whose goal is to split the graph nodes into disjoint $k$ parts in such a way that most edges will connect the nodes within the parts and the number of edges that span two parts is minimized \cite{bulucc2016recent}. The modularity optimization is one of them. The modularity metric measures how far is the number of edges that appear within the parts from that in the random graph model \cite{erdos2011evolution} with the same number of nodes and expected number of edges. This metric has been confirmed to reflect the properties of community existence in many applications. Intuitively, when the assignment of nodes to partitions produces a large modularity, the partitions are likely to be real communities in many different applications including social networks, biological and engineered systems.

For a formal definition, let $G = (V, E)$ be an undirected simple graph with $|V| = n$ nodes and $|E| = m$ edges. The adjacency matrix of $G$ is denoted by $A = \{A_{u,v}\}_{1\leq u,v \leq n}$, where $A_{u,v} = 1$ if there is an edge between node $u$ and node $v$, and 0 otherwise. The degree of a node $v$ is denoted by $d_v$. A $k$-community clustering $\mathcal{C} = \{C_1, \cdots, C_k\}$ is a partition of $V$ into $k$ disjoint sets, namely, $\bigcup_{i=1}^k C_i = V$, and $C_i \bigcap C_j = \varnothing$ for all $1 \leq i \not= j \leq k$. Furthermore, $c_v$ denotes the membership of node $v$ for a given clustering; that is, if $v \in C_i$, then $c_v = i$. The modularity of a clustering $\mathcal{C}$ is given by: \begin{equation}\label{eq:modularity}
\mathcal{Q}(\mathcal{C}) = \frac{1}{2m} \sum_{u, v=1}^{n}B_{u, v} \delta(c_u, c_v),
\end{equation}
where the modularity matrix $B$ is given by $B_{u,v} = A_{u,v} -\frac{d_ud_v}{2m}, 1\leq u,v \leq n$, and $\delta$ is the Kronecker delta: \begin{equation}\label{eq:delta}
\delta(c_u, c_v) = \begin{cases}
    1, & \text{if $c_u = c_v$}\\
    0, & \text{otherwise.}
  \end{cases}
\end{equation}
Our goal is to find the clustering $\mathcal{C}^*$ such that the modularity is maximized:
\begin{equation*}
\mathcal{C}^* = \operatorname{argmax}_{\mathcal{C}} \mathcal{Q}(\mathcal{C}).
\end{equation*}

The problem has applications in chemistry~\cite{niklasson2016graph}, biology~\cite{jeong2000large}, social sciences~\cite{ugander2011anatomy}, and other fields. The task of solving the modularity maximization problem to optimality is NP-complete~\cite{brandes2006maximizing}.

Community detection has been extensively studied classically~\cite{nascimento2011spectral, newman2006modularity}, as well as by using the D-Wave quantum annealer~\cite{negre2020detecting, shaydulin2019network, ushijima2017graph, ushijima2019multilevel} and QAOA~\cite{shaydulin2018community,shaydulin2019network, ushijima2019multilevel}. In these hybrid quantum-classical approaches, the optimization problem is encoded as an Ising model Hamiltonian that has only two-body terms. In the formulations, for a graph with $n$ nodes, solving the 2-community modularity maximization problem requires $n$ qubits, where each qubit encodes the membership of a node. For the $k$-community problem, to encode the membership of each node, one will need to associate $k$ qubits to each node, while introducing quadratic penalty constraints into the Ising Hamiltonian to enforce that each node  belongs to only one community. The formulation requires $kn$ qubits.

\section{Related Work}
\label{sec:related}

The question of ansatz choice is central to the success of hybrid quantum-classical methods introduced in Section \ref{sec:backgroundCOonquantum}. In VQE, the choice of the ansatz determines the expressivity of the trial state and the hardness of finding parameters; therefore, the quality of VQE is only as good as the ansatz. Different strategies of parameterizing the ansatz and updating the parameters will also affect the performance of the algorithm. While being able to reach any state requires a circuit with exponential depth, shallow circuits are preferred in applications, especially if the goal is to run the circuits on modern NISQ devices. McClean et al.~\cite{mcclean2018barren} show that with random parameterized circuit initialization, the exponential dimension of the Hilbert space and the gradient estimation complexity make the optimization impossible for deep circuits. Moreover, Wang et al.~\cite{wang2020noise} show that another type of ``barren plateau'' is induced by hardware noise. More specifically, given local Pauli noise, the gradient vanishes exponentially with the depth of the circuits. Similar results have been demonstrated for QAOA~\cite{xue2019effects}.

There are fundamentally two ways to approach the problem of designing compact ans\"atze. We classify ans\"atze into two groups. The first way is to start with physics or chemistry-inspired ans\"atze generated by a Hamiltonian. These circuits in the first group are typically high depth.  For example, in quantum chemistry, we would use the unitary coupled cluster method ansatz. In particular, unitary coupled cluster with singles and doubles (UCCSD)~\cite{bartlett1989alternative} ansatz can be used in VQE simulations~\cite{peruzzo2014variational}. Each parameter in UCCSD ansatz parametrizes a coupled cluster amplitude for each fermionic excitation from a reference state, either single or double. While it is an accurate ansatz, it has many redundant and unimportant parameters corresponding to the excited states not contributing to the ground energy giving a lot of room for optimization. One strong idea is to use symmetry to generate compact circuits~\cite{gard2020efficient} and another is to use MP2 amplitude to screen out UCCSD parameters~\cite{romero2018strategies}. The issue is that there is a limit to the reduction and typically such circuits are still too deep to execute on NISQ devices, especially for large and moderate size problems.
In the second group are so-called ``hardware-efficient'' (HE) ans\"atze. These ans\"atze contain sequences of parametrized single-qubit and two-qubit gates that can be easily implemented on NISQ devices because of their by design compact nature~\cite{kandala2017hardware,kandala2019error, Barkoutsos_2018,Ganzhorn2019,gard2020efficient}. The key aspect is that no information about the physics or chemistry of the system is used. The major plus of HE ans\"atze is that they are very expressible and the circuits to implement them are shallower compared to the first group of ans\"atze like UCCSD and contain a much smaller number of two-qubit gates. The downside of HE ans\"atze is that they can have too many parameters to optimize and suffer from the barren plateaus~\cite{mcclean2018barren,wang2020noise,cerezo20cost} problem, which was discussed earlier in this paper.

Another way to build ans\"atze is to dynamically generate them using some criteria (for example, using largest gradients or resolutions of identity as a criterion to minimize total energy)~\cite{grimsley2019adaptive,Tang_qubit_adapt,qcc,it_QCC,lang2020unitary,humble_benchmarking,sim2021adaptive}. We will refer to these ans\"atze as ``iterative ans\"atze''. In quantum chemistry, one of the first proposed iterative ans\"atze is ADAPT-VQE~\cite{grimsley2019adaptive}, where fermionic operators are added to the ansatz based on the energy gradients with respect to variational parameters. Later, qubit-ADAPT-VQE~\cite{Tang_qubit_adapt} method was invented, where the fermionic operators are broken down into Pauli strings and used as building blocks for constructing an ansatz. Other important contributions to this field were made by on Yordanov et al.~\cite{qubit_exc} work to use so-called qubit-excitations~\cite{iterative_qibit_exc} instead of the fermionic excitations in ADAPT-VQE simulations and qubit coupled cluster (QCC)~\cite{qcc} method and its iterative version~\cite{it_QCC}, where the ansatz is constructed directly in the qubit space. In general, the iterative circuits are somewhere in the middle between Hamiltonian ans\"atze and HE ans\"atze in terms of depth and the number of parameters.
In the same spirit, Zhu et al. propose an adaptive version of QAOA, called ADAPT-QAOA~\cite{zhu2020adaptive}. Compared with the standard QAOA ansatz, which alternates between the predefined exponentiated cost and mixing Hamiltonian operator, ADAPT-QAOA grows the ansatz with two operators at a time. It also uses a gradient criterion to select the mixing operator from a predefined operator pool. On a class of MaxCut graph problems, ADAPT-QAOA demonstrates faster convergence while also reducing the number of optimization parameters and the \cnotgate~gate counts, compared with standard QAOA.
Our L-VQE approach can be considered as a new evolution of iterative ans\"atze. The traditional iterative ans\"atze are still based on using insights from the physics and chemistry of the problem with some criterion to build ans\"atze. In L-VQE, HE ideas are used to grow circuits. As a result, our circuits are more compact than Hamiltonian or iterative ans\"atze.

The optimization of the parameters is also a key component in hybrid quantum-classical algorithms. For example, in quantum machine learning, Skolik et al.~\cite{skolik2020layerwise} propose a layer-wise learning strategy that grows the circuit depth incrementally during optimization and only updates subsets of parameters in training. However, a recent paper~\cite{campos2020abrupt} shows that this type of layer-wise training strategy, namely, training a circuit piecewise in sequence, could encounter abrupt transitions in the training process as the depth of the circuit grows.

\section{Layer VQE}\label{sec:lvqe}

We advocate an iterative hybrid approach to quantum optimization on NISQ devices, which we call Layer VQE (L-VQE). L-VQE combines ideas from recent developments in adaptive variational algorithms, such as~\cite{grimsley2019adaptive,zhu2020adaptive,skolik2020layerwise}. In this section, we describe L-VQE in detail.

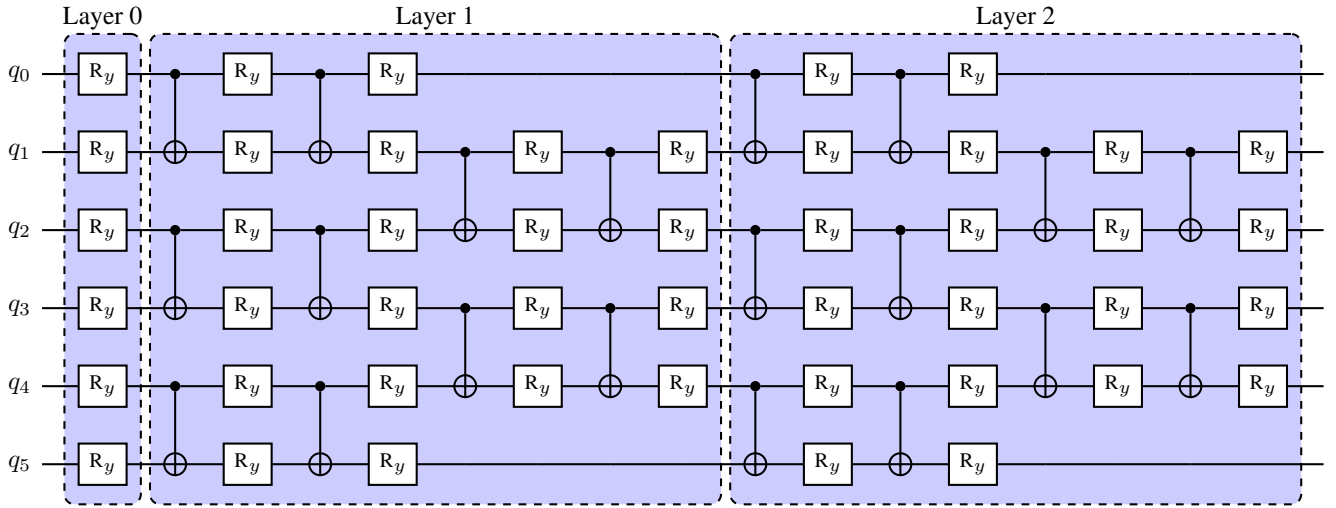
\begin{figure*}[htbp!]
\resizebox{\textwidth}{!}{
\begin{quantikz}
\lstick{$q_0$} & \gate{\rygate} \gategroup[6,steps=1,style={dashed,
rounded corners,fill=blue!20,inner xsep=2pt},
background]{{Layer 0}} & \ctrl{1} \gategroup[6,steps=8,style={dashed,
rounded corners,fill=blue!20,inner xsep=2pt},
background]{{Layer 1}} & \gate{\rygate} & \ctrl{1} & \gate{\rygate} & \qw & \qw & \qw & \qw & \ctrl{1} \gategroup[6,steps=8,style={dashed,
rounded corners,fill=blue!20,inner xsep=2pt},
background]{{Layer 2}} & \gate{\rygate} & \ctrl{1} & \gate{\rygate} & \qw & \qw & \qw & \qw & \qw \\
\lstick{$q_1$} & \gate{\rygate} & \targ{}  & \gate{\rygate} & \targ{}  & \gate{\rygate} & \ctrl{1} & \gate{\rygate} & \ctrl{1} & \gate{\rygate} & \targ{}  & \gate{\rygate} & \targ{}  & \gate{\rygate} & \ctrl{1} & \gate{\rygate} & \ctrl{1} & \gate{\rygate} & \qw \\
\lstick{$q_2$} & \gate{\rygate} & \ctrl{1} & \gate{\rygate} & \ctrl{1} & \gate{\rygate} & \targ{}  & \gate{\rygate} & \targ{}  & \gate{\rygate} & \ctrl{1} & \gate{\rygate} & \ctrl{1} & \gate{\rygate} & \targ{}  & \gate{\rygate} & \targ{}  & \gate{\rygate} & \qw \\
\lstick{$q_3$} & \gate{\rygate} & \targ{}  & \gate{\rygate} & \targ{}  & \gate{\rygate} & \ctrl{1} & \gate{\rygate} & \ctrl{1} & \gate{\rygate} & \targ{}  & \gate{\rygate} & \targ{}  & \gate{\rygate} & \ctrl{1} & \gate{\rygate} & \ctrl{1} & \gate{\rygate} & \qw \\
\lstick{$q_4$} & \gate{\rygate} & \ctrl{1} & \gate{\rygate} & \ctrl{1} & \gate{\rygate} & \targ{}  & \gate{\rygate} & \targ{}  & \gate{\rygate} & \ctrl{1} & \gate{\rygate} & \ctrl{1} & \gate{\rygate} & \targ{}  & \gate{\rygate} & \targ{}  & \gate{\rygate} & \qw \\
\lstick{$q_5$} & \gate{\rygate} & \targ{}  & \gate{\rygate} & \targ{}  & \gate{\rygate} & \qw      & \qw        & \qw      & \qw        & \targ{}  & \gate{\rygate} & \targ{}  & \gate{\rygate} & \qw      & \qw       & \qw      & \qw        & \qw
\end{quantikz}}
\caption{L-VQE ansatz for a 6-qubit quantum state. $\rygate$ denotes rotation around the $y$-axis defined as $\rygate(\theta) \equiv e^{-i\frac{\theta}{2}\ygate}$. Every $\rygate$ contains a parameter that is optimized over in the outer loop.}
\label{fig:circuit1}
\end{figure*}

Suppose we use a problem encoding that requires $n$ qubits. We start the algorithm with an ansatz with no entangling gates and one $\rygate$ gate acting on each qubit, where $\rygate$ is the single qubit rotation through an angle $\theta$ around the $y$-axis, the unitary matrix is defined as $\rygate(\theta) \equiv e^{-i\frac{\theta}{2}\ygate}$, and $\ygate$ is the Pauli $\ygate$ operator.
The parameters of these $\rygate$ gates are initialized uniformly randomly on $[0, 2\pi]$. We denote the parameters for this layer of gates (\emph{Layer 0} in Fig. \ref{fig:circuit1}) as $\bm{\theta}_0$ and the layer as $U_0(\bm{\theta}_0)$. The quantum state after applying the circuit to the initial state $\ket{0}$ is denoted as $\ket{\psi_0(\bm{\theta}_0)} \equiv U_0(\bm{\theta}_0)\ket{0}$. We then proceed to the conventional VQE routine and iteratively update the parameters $\bm{\theta}_0$ to minimize the cost function $\braket{\psi_0(\bm{\theta}_0) | \mathcal{H} | \psi_0(\bm{\theta}_0)}$. In conventional VQE, this iterative procedure is run until convergence; but in L-VQE, we stop after a fixed number of iterations and then add another set of gates to the ansatz. The conventional strategy can indeed produce a better result at this step, but after adding the new set of gates, it may more easily get trapped in a local minimum in the subsequent optimization procedure. In our experiments, the number of iterations is picked empirically and increases linearly as system size grows.

The newly added set of gates includes the $\rygate$ gates and $\cnotgate$ gates that act on nearest-neighbor qubits. Another way to describe this whole procedure is that we embed the obtained parameterized circuit into a deeper circuit. We denote this newly added layer of the circuit $U_1(\bm{\theta}_1)$, (\emph{Layer 1} in Fig. \ref{fig:circuit1}). The newly added parameters $\bm{\theta}_1$ are initialized as zero. Note that here since $\rygate(0) = \idgate$ and $\cnotgate^2 = \idgate$, where $\idgate$ is the identity matrix, the quantum state becomes
\begin{equation}
\ket{\psi_1(\bm{\theta}_0, \bm{\theta}_1)} \equiv U_1(\bm{\theta}_1) U_0(\bm{\theta}_0)\ket{0} = U_0(\bm{\theta}_0)\ket{0} = \ket{\psi_0(\bm{\theta}_0)}.
\end{equation}

Therefore, initializing the newly added parameters as zeros guarantee that the cost function that we are optimizing will not change after adding this new layer. At this point, we can either let the optimization run until convergence or repeat the previous process, stop at a fixed number of iterations, and add another set of gates to the circuit and then optimize. The pseudo code of the algorithm is presented in Algorithm \ref{alg:lvqe}.

\begin{algorithm}[H]
\begin{algorithmic}[1]
\caption{L-VQE with $\ell$ layers}
\label{alg:lvqe}
\STATE Initialize the ansatz with one $\rygate$ acting on each qubit.
\STATE Update the parameters to minimize $\braket{\psi_0(\bm{\theta}) | \mathcal{H} | \psi_0(\bm{\theta})}$; stop after $k_0$ iterations (before reaching convergence).
\FOR{$l = 1, \cdots, \ell$}
\STATE Add a new layer to the ansatz, and initialize it such that it evaluates to identity.
\STATE Update all parameters to minimize $\braket{\psi_{\ell}(\bm{\theta}) | \mathcal{H} | \psi_{\ell}(\bm{\theta})}$; stop after $k_{\ell}$ iterations (before reaching convergence).
\ENDFOR
\STATE Update all parameters to minimize $\braket{\psi(\bm{\theta}) | \mathcal{H} | \psi(\bm{\theta})}$ until convergence.
\end{algorithmic}
\end{algorithm}

In simulations, the cost function $\braket{\psi(\bm{\theta}) | \mathcal{H} | \psi(\bm{\theta})}$ can be evaluated exactly. When executing the algorithm on hardware, we have to repeat the state preparation and measurement  multiple times to generate a number of samples, and use the samples to estimate the cost function, introducing an error due to the finite number of samples. In our experiments we investigate the performance of the algorithm in both cases.

The solution of a combinatorial optimization problem is classical, i.e., it is a computational basis state. Therefore, suppose we can find optimal parameters for the ansatz; the ansatz contains only single qubit rotations on each qubit, should be able to prepare the state that contains the computational basis state we want. However, ansatz as such is prone to local minima. Therefore in L-VQE, we start from an ansatz that only contains single qubit rotation gates, then iteratively add entanglement to the ansatz to help the optimization process. Another motivation is to speed up the optimization process. Random initialization of deep circuit leads to difficult optimization, so we ``pre-train'' with shallower circuits. Importantly, we do not converge the optimization for a single layer (or other shallow circuits). Indeed, this would cause the algorithm to get stuck in a local minimum, which might be hard to escape after a new layer is added. Instead, we add a new layer before convergence is reached to avoid the above-mentioned issues. Empirical results show that this strategy increases the probability of finding the ground state or finding the state that is sufficiently close to the ground state.

Similar to ADAPT-VQE~\cite{grimsley2019adaptive} and ADAPT-QAOA~\cite{zhu2020adaptive}, we grow the size of the ansatz as we iteratively update the parameters. The added parameterized ansatz is initialized such that the new circuit parts evaluate to identity in order to avoid deterioration of the optimization. In ADAPT-VQE and ADAPT-QAOA, however, the algorithm will identify an operator that has the largest gradient from a collection of operators and then add this operator to the ansatz. In L-VQE, we define the newly added ansatz upfront.

As discussed in~\cite{campos2020abrupt}, the conjecture that a circuit can be trained piece-wise turns out to not always be true. In the finite setting, there are abrupt transitions in the ability of quantum circuits to be trained. In layer-wise learning~\cite{skolik2020layerwise}, when adding a new set of layers, part of the previous layer's parameters are frozen, and additional optimization sweeps are performed on subsets of parameters. Each layer contains rotation gates on each qubit and also operators that connect the qubits. L-VQE, on the other hand, optimizes all parameters where none of the previous layers are fixed. Furthermore, the initial ansatz for L-VQE only contains rotation gates on each qubit. Therefore, we start from a product state, where no entanglement is involved. After some iterations, we add entanglement to help the optimization process. This is different from \cite{skolik2020layerwise}, where the initial layer already contains operators connecting qubits. This strategy may reduce the limitations of the lack of layer-wise trainability \cite{campos2020abrupt}. In addition, layer-wise learning is a general approach; for L-VQE, we focus on solving combinatorial optimization problems.

\section{The $k$-Community Detection}\label{sec:modularity}

We propose a novel qubit-frugal formulation for the $k-$community detection problem.  When the problem is to divide the network into two communities, namely, with $k=2$, we can associate a binary variable with each node $v \in V$ such that \begin{equation*}
x_v = \begin{cases}
    1, & \text{if $c_v = 1$}\\
    0, & \text{if $c_v = 2$.}
  \end{cases}
\end{equation*}

Then, we can rewrite the Kronecker delta (\ref{eq:delta}) in terms of these binary variables: \begin{equation}\label{eq:delta1}
\delta(c_u, c_v) = \delta(x_u, x_v) = 2x_ux_v - x_u - x_v + 1.
\end{equation}

Plugging (\ref{eq:delta1}) into (\ref{eq:modularity}) leads to the expression of modularity: \begin{equation*}
\mathcal{Q}(\mathcal{C}) = \frac{1}{2m} \sum_{u, v=1}^{n}B_{u, v} (2x_ux_v - x_u - x_v + 1).
\end{equation*}

For larger $k$, we can use a binary encoding by associating $N = \lceil \log_2 k\rceil$ binary variables $\{x_{j, v}\}_{j=1}^N \subset \{0, 1\}^N$ with each node $v \in V$. We can rewrite the membership of node $v$ as \begin{equation*}
c_v = \sum_{j=1}^{N} 2^{j-1} x_{j,v}.
\end{equation*}

Again, we can rewrite the Kronecker delta (\ref{eq:delta}) in terms of these binary variables: \begin{equation}\label{eq:delta2}
\delta(c_u, c_v) = \prod_{j=1}^N \delta(x_{j, u}, x_{j, v}) = \prod_{j=1}^N (2x_{j, u}x_{j, v} - x_{j, u} - x_{j, v} + 1).
\end{equation}

Plugging (\ref{eq:delta2}) into (\ref{eq:modularity}), we obtain for the modularity \begin{equation}\label{eq:modularity2}
\mathcal{Q}(\mathcal{C}) = \frac{1}{2m} \sum_{u, v=1}^{n}B_{u, v} \prod_{j=1}^N  (2x_{j, u}x_{j, v} - x_{j, u} - x_{j, v} + 1).
\end{equation}

Following the construction described in Section \ref{sec:background}, maximizing the modularity in (\ref{eq:modularity2}) can be formulated in terms of finding the ground state of the following Hamiltonian, \begin{equation}\label{eq:hamiltonian}
\mathcal{H} = - \frac{1}{2m} \sum_{u, v=1}^{n}B_{u, v} \prod_{j=1}^N \frac{I + \zgate_{j, u} \zgate_{j, v}}{2},
\end{equation}
 where binary variables $x_{j, v}$ have been substituted with $\frac{1}{2}(I - \zgate_{j, v}), \forall j \in \{1, 2, \cdots, N\}, \forall v \in V$. Here, $\zgate_{j, v}$ is the Pauli $\zgate$ operator that acts on qubit $(j, v)$.

Other formulations have been proposed to tackle the problem for specific quantum architectures. Ushijima-Mwesigwa et al.~\cite{ushijima2017graph} use an Ising Hamiltonian formulation to detect two communities using quantum annealing on the D-Wave system, which requires $n$ qubits. Negre et al.~\cite{negre2020detecting} extend it to detect $k$ communities, which requires $kn$ qubits. In contrast, the Hamiltonian we propose in this work requires only $n \left\lceil \log_2 k \right\rceil$ qubits thanks to the encoding introduced above. Note that the many-body interactions present in the proposed Hamiltonian do not introduce significant overhead, as simulating a product of $N$ Pauli $\zgate$ operators requires only $2(N-1)$ $\cnotgate$s.

\section{Experiments}\label{sec:experiments}

In this section we present the numerical results. Since QAOA is considered the leading approach for combinatorial optimization on NISQ devices, we begin in Section \ref{sec:experimentsqaoa} with a numerical comparison of L-VQE and QAOA. We then compare L-VQE with the second leading approach, which is VQE in Section \ref{sec:experimentsvqe}. To highlight the potential of the proposed L-VQE approach on NISQ devices, we present some further evidence in Section \ref{sec:experimentsevidence}. This includes a scalability analysis and simulation results of L-VQE on a trapped ion noisy quantum simulator with a realistic noise level. To highlight the importance of entanglement for optimization, in Section~\ref{sec:experimentsentangle} we present results comparing VQE with and without entanglement.

\subsection{L-VQE and QAOA} \label{sec:experimentsqaoa}

For the first set of experiments, we run simulations for the L-VQE and QAOA algorithm with the proposed Hamiltonian (\ref{eq:hamiltonian}). The goal is to find a clustering of up to 4 communities that maximize the modularity. We are thus simulating $2n$ qubits for a graph with $n$ nodes. For L-VQE, we run our simulations of the quantum circuits in MATLAB. We use matrix product states (MPS) techniques to simulate quantum circuits, which allows us to reach large system sizes (up to 40 qubits and 352 parameters). The Hamiltonian is also represented in the form of a matrix product operator~\cite{orus2014practical}, allowing access to full precision energy computation. Our proprietary MPS simulator uses an exact representation of the wave function without any truncation. The complexity of simulations in the MPS simulator scales linearly with system size and exponentially with circuit depth. Since the ansatz in L-VQE is 1-dimensional and shallow, the MPS simulator can simulate L-VQE circuits for relatively large system sizes. The QAOA circuits we consider are deep; therefore there is no benefit to using the MPS simulator. We use the high-performance simulator Qiskit Aer~\cite{Qiskit} to simulate QAOA circuits due to its convenience. Because of the simulation complexity and the need to optimize parameters for the benchmark instances, we limit the simulations of QAOA to 20 qubits. The choice of simulator (MPS or Qiskit Aer) is inconsequential, as both methods simulate the quantum state exactly and produce the same outcomes.

In variational algorithms, the choice of the classical outer-loop optimizer is central to the performance of the method. However, in this work, we do not specifically investigate the performance of various optimizers and do not perform any hyperparameter tuning on the optimizers. We only test L-VQE with a sequential minimal optimizer (SMO)~\cite{nakanishi2019sequential} and COBYLA~\cite{powell1994direct,powell1998direct}. SMO is implemented using the recommended settings~\cite{nakanishi2019sequential}, and COBYLA is implemented in the SciPy~\cite{scipy} package with the default setting. We observe that SMO performs slightly better than COBYLA, therefore we advocate for SMO over COBYLA. For optimization in QAOA, we also use COBYLA with the default setting (we do not consider SMO as it is not designed for QAOA). Furthermore, we exhaustively optimized the parameters by using COBYLA as a local optimizer in the libEnsemble~\cite{libEnsemble_0.5.0} implementation of APOSMM~\cite{LarWild14,LW16}. Given a fixed number of iterations, APOSMM, as a multistart method, will run the local optimizer until convergence and then restart the optimization. This approach has been shown to work well in our previous study~\cite{shaydulin2019multistart}. Additional details of the QAOA experiments are provided in Appendix \ref{app:qaoa}.

\begin{figure}[!htbp]
\centering
\includegraphics[width=\linewidth]{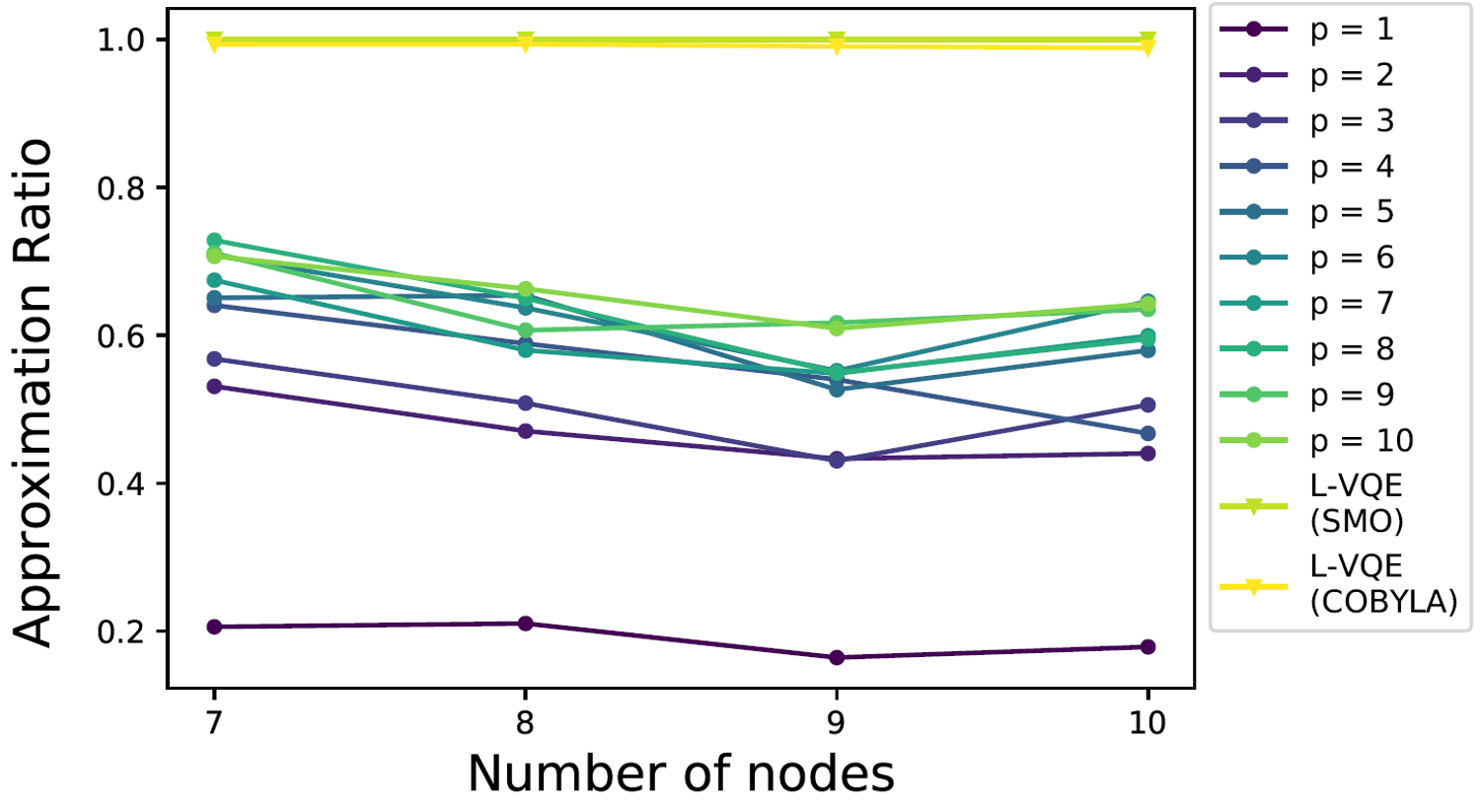}
\caption{Compare QAOA and L-VQE on graphs of size from 7 to 10, simulating 14--20 qubits. L-VQE ansatz is iteratively increased up to $\ell=2$ layers. L-VQE finds the ground state or a state that is close (with approximation ratio at least 0.99) to the ground state for each graph.}
\label{fig:qaoascale7}
\end{figure}

We run QAOA experiments on 4 \texttt{gnp random} graphs, with 7, 8, 9, 10 nodes respectively, and with $p$ up to 10. All graphs are generated with \texttt{Networkx}. We give APOSMM a limit of 30,000 iterations. The limit is chosen based on an empirical observation that with this parameter choice APOSMM will restart COBYLA for at least 10 times, usually much more. To compare, we run our L-VQE on each graph 10 times given different random seed using SMO and COBYLA as the optimizer. Each run is given a limit of 3,000 iterations, and we report the best result found by L-VQE. The results of the experiments are shown in Fig. \ref{fig:qaoascale7}. For each graph, L-VQE finds an estimate of the ground state with an approximation ratio $\rho$ of at least 0.99, despite having a much lower budget of function evaluation. 


For a graph with $n$ nodes, our approach requires $2n$ qubits in order to detect 4 communities. Assuming full connectivity and compiling the higher-order terms in the Hamiltonian (\ref{eq:hamiltonian}) into gate sets $\{\rzgate, \cnotgate\}$, the product of 4 Pauli $\zgate$ operators is decomposed into $6$ $\cnotgate$s and one $\rzgate$ in the middle. Single-qubit gates are cheap, so one is primarily interested in the $\cnotgate$ gate counts. The gate counts of QAOA and L-VQE circuits are summarized in Table \ref{tab:gatecount}. When $p$ is small, the Hamiltonian evolution ansatz used in QAOA is less expressive as compared with the hardware-efficient ansatz used in L-VQE. Therefore a large number of QAOA layers and large circuit depth is needed to achieve the required overlap with the target state. At the same time, the cost function landscape of QAOA is highly nonconvex and contains many low-quality local optima, which make finding high-quality parameters difficult for larger $p$. In addition, the Hamiltonian (\ref{eq:hamiltonian}) contains many-body terms, which can be hard to compile into gates in practice due to the limited connectivity of the hardware.
In contrast, L-VQE follows the connectivity of the hardware, as the Hamiltonian structure does not enter the ansatz explicitly.

\begin{table}
\centering
\caption{Assuming full connectivity and compiling the higher-order terms in the Hamiltonian (\ref{eq:hamiltonian}) into gate sets $\{\rzgate, \cnotgate\}$, the gate count of QAOA scales quadratically with $n$, while L-VQE scales linearly. In our experiments presented in Fig. \ref{fig:qaoascale7_1}, QAOA circuits with $p$ steps consists of $77p$ single qubit gates and 210$p$ $\cnotgate$ gates, while L-VQE with $\ell$ layers contains $52\ell + 14$ single qubit gates and $26\ell$ $\cnotgate$ gates. Thus, we expect that the L-VQE approach will be more robust to noise in real-life experiments. $\cnotgate$ count of QAOA can be decreased by further circuit optimizations and more efficient native gates. On the other hand, it would be increased if the connectivity is not full.}
\label{tab:gatecount}
\begin{tabular}{|c|c|c|c|}
\hline
 & QAOA with $p$ steps & L-VQE with $\ell$ layers \\
\hline
$\cnotgate$ & $5n(n-1)p$ & $\ell(4n-2)$ \\
\hline
single qubit gates & $\frac{1}{2}(3n^2+n)p$ & $\ell(8n-4)+2n$ \\
\hline
\end{tabular}
\end{table}

\subsection{VQE and L-VQE}\label{sec:experimentsvqe}

To further examine the performance of L-VQE, we compare the results of VQE with fixed ansatz and L-VQE on larger problems. In Section \ref{sec:experimentsvqesample} we compare the performance of VQE and L-VQE, and in Section \ref{sec:experimentsvqeenergy} we compare the performance with full precision energy computation, that is with energy computed as the expectation of the problem Hamiltonian with the full simulated quantum state. The  results are summarized in Section~\ref{sec:experimentsvqesummary}.

We generated 16 graph instances with \texttt{NetworkX}. Graph information is summarized in Table \ref{tab:graphinfo}. The goal is to find a clustering of up to 4 communities that maximizes the modularity; thus we are simulating 34 qubits for \texttt{windmill} and 40 qubits for all other graphs.

\begin{table}
\centering
\caption{Graph information of the \texttt{Networkx} generated instances for comparing VQE and L-VQE.}\label{tab:graphinfo}
\begin{tabular}{|c|c|c|}
\hline
Graph class & \# instances & $|V|$ \\
\hline
\texttt{relaxed caveman} & 2 & 20 \\
\hline
\texttt{gaussian random} & 2 & 20 \\
\hline
\texttt{random partition} & 4 & 20 \\
\hline
\texttt{windmill} & 1 & 17 \\
\hline
\texttt{gnp random} & 4 & 20 \\
\hline
\texttt{power law cluster} & 3 & 20 \\
\hline
\end{tabular}
\end{table}

For VQE, we define a fixed form of the ansatz upfront and then iteratively optimize and update over all parameters. We compare 3 sets of ans\"atze, which are shown in Fig. \ref{fig:circuit1} as Layer 0 only ($\ell=0$), Layer 0 to 1 only ($\ell=1$), and Layer 0 to 2 ($\ell=2$), respectively. For L-VQE with $\ell=0$, the ansatz will not grow; thus the algorithm is the same as VQE with  one $\rygate$ gate acting on each qubit. For L-VQE with $\ell=1$ and $\ell=2$, we set the parameter $k_0 = 200$ in Algorithm \ref{alg:lvqe}. In other words, we first run L-VQE with Layer 0 ansatz for 200 iterations and then reuse the parameters to the ansatz with 1 layer and 2 layers, respectively. Again, we run our simulations of the quantum circuits in the MPS simulator. For optimization, we use the sequential minimal optimizer~\cite{nakanishi2019sequential} and COBYLA~\cite{powell1994direct}. For each graph and each approach, we initialize the ansatz with 10 different random seeds.

\subsubsection{VQE and L-VQE}
\label{sec:experimentsvqesample}

We report the results of VQE and L-VQE in Table \ref{tab:bestnoiseSMO}--\ref{tab:averagenoiseCOBYLA}. To evaluate the cost function $\braket{\psi(\bm{\theta}) | \mathcal{H} | \psi(\bm{\theta})}$, we execute the circuit and generate 2,000 samples and use the mean of the samples as an estimator. Having a finite number of samples is a realistic setup, since when the scale of the system gets larger, the exact computation of the cost function becomes intractable.

In Table \ref{tab:bestnoiseSMO}, we report the best approximation ratio ($\rho_{\text{best}}$) achieved from the 10 runs using SMO for each graph. In Table \ref{tab:averagenoiseSMO}, we report the average and standard deviation ($\rho_{\text{average}} \pm \sigma$) of the approximation ratio from the 10 runs for each graph. We additionally report the results that use COBYLA as the optimizer in Tables \ref{tab:bestnoiseCOBYLA}, \ref{tab:averagenoiseCOBYLA}.

\begin{table}
\centering
\caption{Best approximation ratio achieved by VQE and L-VQE using SMO. As the number of layers in the ansatz increases, results of VQE deteriorates. L-VQE does not suffer from that problem, and we achieve better results as the number of layers grows.}\label{tab:bestnoiseSMO}
\begin{tabular}{|c|c|c|c|c|c|}
\hline
\multirow{2}{*}{graph} & \multicolumn{2}{c|}{VQE $\rho_{\text{best}}$} & \multicolumn{3}{c|}{L-VQE $\rho_{\text{best}}$} \\ \cline{2-6}
      & 1 Layer & 2 Layer & 0 Layer & 1 Layer & 2 Layer \\
\hline
caveman & 0.99 & 0.93 & 1.0 & 1.0 & 1.0 \\\hline
caveman2 & 0.98 & 0.93 & 0.99 & 0.99 & 0.99 \\\hline
gaussian & 0.98 & 0.9 & 1.0 & 1.0 & 1.0 \\\hline
gaussian2 & 0.99 & 0.93 & 0.99 & 1.0 & 1.0 \\\hline
gnp1 & 0.99 & 0.93 & 1.0 & 1.0 & 1.0 \\\hline
gnp2 & 0.98 & 0.92 & 0.95 & 1.0 & 1.0 \\\hline
gnp3 & 0.99 & 0.91 & 0.99 & 1.0 & 1.0 \\\hline
gnp4 & 0.99 & 0.88 & 0.97 & 0.99 & 1.0 \\\hline
power & 0.96 & 0.9 & 1.0 & 1.0 & 1.0 \\\hline
power2 & 0.99 & 0.93 & 0.99 & 1.0 & 1.0 \\\hline
power3 & 0.98 & 0.91 & 0.9 & 0.96 & 1.0 \\\hline
random1 & 0.98 & 0.93 & 1.0 & 1.0 & 1.0 \\\hline
random2 & 0.99 & 0.89 & 0.95 & 1.0 & 1.0 \\\hline
random3 & 0.99 & 0.96 & 1.0 & 1.0 & 1.0 \\\hline
random4 & 0.92 & 0.93 & 0.96 & 1.0 & 0.98 \\\hline
windmill & 0.99 & 0.96 & 1.0 & 1.0 & 1.0 \\\hline
\end{tabular}
\end{table}

\begin{table*}
\centering
\caption{Average approximation ratio achieved by VQE and L-VQE using SMO. As the number of layers in the ansatz increases, results of VQE deteriorate; but for L-VQE, we achieve better results.}\label{tab:averagenoiseSMO}
\begin{tabular}{|c|c|c|c|c|c|}
\hline
\multirow{2}{*}{graph} & \multicolumn{2}{c|}{VQE $\rho_{\text{average}} \pm \sigma$} & \multicolumn{3}{c|}{L-VQE $\rho_{\text{average}} \pm \sigma$} \\
\cline{2-6}
      & 1 Layer & 2 Layer & 0 Layer & 1 Layer & 2 Layer \\
\hline
caveman & 0.91 $\pm$ 0.08 & 0.83 $\pm$ 0.08 & 0.83 $\pm$ 0.15 &  0.92 $\pm$ 0.1 & 0.91 $\pm$ 0.1 \\\hline
caveman2 & 0.95 $\pm$ 0.03 & 0.86 $\pm$ 0.06 & 0.92 $\pm$ 0.07 &  0.99 $\pm$ 0.0 &  0.99 $\pm$ 0.0 \\\hline
gaussian & 0.83 $\pm$ 0.08 & 0.77 $\pm$ 0.09 & 0.86 $\pm$ 0.08 & 0.94 $\pm$ 0.09 & 0.91 $\pm$ 0.09 \\\hline
gaussian2 & 0.92 $\pm$ 0.06 & 0.87 $\pm$ 0.03 & 0.9 $\pm$ 0.07 &  0.95 $\pm$ 0.06 &  0.95 $\pm$ 0.05 \\\hline
gnp1 & 0.87 $\pm$ 0.07 & 0.82 $\pm$ 0.08 & 0.87 $\pm$ 0.07 & 0.93 $\pm$ 0.05 &  0.94 $\pm$ 0.06 \\\hline
gnp2 & 0.89 $\pm$ 0.07 & 0.8 $\pm$ 0.09 & 0.87 $\pm$ 0.07 & 0.92 $\pm$ 0.06 &  0.94 $\pm$ 0.04 \\\hline
gnp3 & 0.89 $\pm$ 0.06 & 0.81 $\pm$ 0.06 & 0.86 $\pm$ 0.07 & 0.94 $\pm$ 0.06 &  0.95 $\pm$ 0.04 \\\hline
gnp4 & 0.92 $\pm$ 0.05 & 0.82 $\pm$ 0.04 & 0.92 $\pm$ 0.05 & 0.95 $\pm$ 0.04 & 0.94 $\pm$ 0.06 \\\hline
power &  0.9 $\pm$ 0.04 & 0.82 $\pm$ 0.05 & 0.87 $\pm$ 0.07 &  0.9 $\pm$ 0.07 & 0.89 $\pm$ 0.08 \\\hline
power2 & 0.93 $\pm$ 0.08 & 0.85 $\pm$ 0.06 & 0.9 $\pm$ 0.06 & 0.92 $\pm$ 0.05 &  0.93 $\pm$ 0.05 \\\hline
power3 & 0.85 $\pm$ 0.07 & 0.79 $\pm$ 0.07 & 0.84 $\pm$ 0.06 & 0.89 $\pm$ 0.04 &  0.9 $\pm$ 0.05 \\\hline
random1 & 0.91 $\pm$ 0.1 & 0.84 $\pm$ 0.05 & 0.86 $\pm$ 0.13 &  0.98 $\pm$ 0.02 &  0.98 $\pm$ 0.02 \\\hline
random2 &  0.93 $\pm$ 0.07 & 0.81 $\pm$ 0.06 & 0.85 $\pm$ 0.1 & 0.93 $\pm$ 0.07 &  0.93 $\pm$ 0.06 \\\hline
random3 & 0.95 $\pm$ 0.05 & 0.84 $\pm$ 0.1 & 0.9 $\pm$ 0.11 & 0.95 $\pm$ 0.04 &  0.97 $\pm$ 0.03 \\\hline
random4 & 0.85 $\pm$ 0.05 & 0.83 $\pm$ 0.07 & 0.82 $\pm$ 0.08 &  0.9 $\pm$ 0.08 &  0.9 $\pm$ 0.07 \\\hline
windmill & 0.93 $\pm$ 0.06 & 0.9 $\pm$ 0.05 & 0.84 $\pm$ 0.12 & 0.92 $\pm$ 0.06 &  0.94 $\pm$ 0.06 \\\hline
\end{tabular}
\end{table*}

\begin{table}
\centering
\caption{Best approximation ratio achieved by VQE and L-VQE using COBYLA. As the number of layers in the ansatz increases, results of VQE deteriorate; but for L-VQE, we achieve better results.}\label{tab:bestnoiseCOBYLA}
\begin{tabular}{|c|c|c|c|c|c|}
\hline
\multirow{2}{*}{graph} & \multicolumn{2}{c|}{VQE $\rho_{\text{best}}$} & \multicolumn{3}{c|}{L-VQE $\rho_{\text{best}}$} \\
\cline{2-6}
      & 1 Layer & 2 Layer & 0 Layer & 1 Layer & 2 Layer \\
\hline
caveman & 0.96 & 0.89 & 0.97 & 1.0 & 1.0 \\\hline
caveman2 & 0.95 & 0.92 & 0.97 & 0.99 & 0.99 \\\hline
gaussian & 0.96 & 0.81 & 0.92 & 1.0 & 1.0 \\\hline
gaussian2 & 0.91 & 0.84 & 0.91 & 1.0 & 1.0 \\\hline
gnp1 & 0.94 & 0.85 & 0.91 & 1.0 & 1.0 \\\hline
gnp2 & 0.86 & 0.84 & 0.87 & 0.97 & 0.97 \\\hline
gnp3 & 0.83 & 0.84 & 0.97 & 1.0 & 1.0 \\\hline
gnp4 & 0.89 & 0.86 & 0.96 & 1.0 & 0.98 \\\hline
power & 0.89 & 0.89 & 0.92 & 0.99 & 0.99 \\\hline
power2 & 0.91 & 0.77 & 0.9 & 1.0 & 1.0 \\\hline
power3 & 0.92 & 0.83 & 0.91 & 1.0 & 1.0 \\\hline
random1 & 0.93 & 0.88 & 0.96 & 1.0 & 1.0 \\\hline
random2 & 0.92 & 0.91 & 0.93 & 1.0 & 1.0 \\\hline
random3 & 0.97 & 0.91 & 0.98 & 1.0 & 1.0 \\\hline
random4 & 0.91 & 0.88 & 0.91 & 1.0 & 1.0 \\\hline
windmill & 0.97 & 0.95 & 0.99 & 1.0 & 1.0 \\\hline
\end{tabular}
\end{table}

\begin{table*}
\centering
\caption{Average approximation ratio achieved by VQE and L-VQE using COBYLA. As the number of layers in the ansatz increases, results of VQE deteriorate, but for L-VQE, we achieve better results.}\label{tab:averagenoiseCOBYLA}
\begin{tabular}{|c|c|c|c|c|c|}
\hline
\multirow{2}{*}{graph} & \multicolumn{2}{c|}{conventional VQE $\rho_{\text{average}} \pm \sigma$} & \multicolumn{3}{c|}{L-VQE $\rho_{\text{average}} \pm \sigma$} \\
\cline{2-6}
      & 1 Layer & 2 Layer & 0 Layer & 1 Layer & 2 Layer \\
\hline
caveman & 0.86 $\pm$ 0.07 & 0.81 $\pm$ 0.06 & 0.8 $\pm$ 0.1 & 0.9 $\pm$ 0.07 & 0.88 $\pm$ 0.06 \\\hline
caveman2 & 0.84 $\pm$ 0.09 & 0.77 $\pm$ 0.12 & 0.82 $\pm$ 0.13 & 0.98 $\pm$ 0.02 & 0.99 $\pm$ 0.0 \\\hline
gaussian & 0.76 $\pm$ 0.15 & 0.63 $\pm$ 0.1 & 0.75 $\pm$ 0.08 & 0.89 $\pm$ 0.08 & 0.87 $\pm$ 0.07 \\\hline
gaussian2 & 0.78 $\pm$ 0.1 & 0.69 $\pm$ 0.09 & 0.8 $\pm$ 0.08 & 0.96 $\pm$ 0.05 & 0.97 $\pm$ 0.04 \\\hline
gnp1 & 0.8 $\pm$ 0.08 & 0.72 $\pm$ 0.09 & 0.79 $\pm$ 0.09 & 0.9 $\pm$ 0.06 & 0.91 $\pm$ 0.07 \\\hline
gnp2 & 0.75 $\pm$ 0.07 & 0.72 $\pm$ 0.1 & 0.78 $\pm$ 0.07 & 0.91 $\pm$ 0.06 & 0.92 $\pm$ 0.05 \\\hline
gnp3 & 0.69 $\pm$ 0.1 & 0.71 $\pm$ 0.1 & 0.79 $\pm$ 0.11 & 0.93 $\pm$ 0.06 & 0.94 $\pm$ 0.07 \\\hline
gnp4 & 0.81 $\pm$ 0.08 & 0.71 $\pm$ 0.11 & 0.85 $\pm$ 0.06 & 0.95 $\pm$ 0.03 & 0.95 $\pm$ 0.03 \\\hline
power & 0.79 $\pm$ 0.07 & 0.67 $\pm$ 0.14 & 0.76 $\pm$ 0.1 & 0.92 $\pm$ 0.05 & 0.92 $\pm$ 0.03 \\\hline
power2 & 0.78 $\pm$ 0.1 & 0.67 $\pm$ 0.08 & 0.78 $\pm$ 0.09 & 0.95 $\pm$ 0.03 & 0.93 $\pm$ 0.07 \\\hline
power3 & 0.77 $\pm$ 0.1 & 0.67 $\pm$ 0.07 & 0.79 $\pm$ 0.09 & 0.9 $\pm$ 0.08 & 0.86 $\pm$ 0.08 \\\hline
random1 & 0.82 $\pm$ 0.09 & 0.74 $\pm$ 0.12 & 0.86 $\pm$ 0.09 & 0.98 $\pm$ 0.03 & 0.97 $\pm$ 0.03 \\\hline
random2 & 0.78 $\pm$ 0.12 & 0.69 $\pm$ 0.15 & 0.82 $\pm$ 0.08 & 0.93 $\pm$ 0.05 & 0.95 $\pm$ 0.04 \\\hline
random3 & 0.84 $\pm$ 0.1 & 0.79 $\pm$ 0.1 & 0.9 $\pm$ 0.06 & 0.94 $\pm$ 0.04 & 0.94 $\pm$ 0.05 \\\hline
random4 & 0.8 $\pm$ 0.09 & 0.71 $\pm$ 0.08 & 0.77 $\pm$ 0.09 & 0.93 $\pm$ 0.06 & 0.94 $\pm$ 0.05 \\\hline
windmill & 0.9 $\pm$ 0.06 & 0.82 $\pm$ 0.09 & 0.87 $\pm$ 0.09 & 0.92 $\pm$ 0.06 & 0.92 $\pm$ 0.06 \\\hline
\end{tabular}
\end{table*}

\subsubsection{VQE and L-VQE with full precision energy computation} \label{sec:experimentsvqeenergy}
We report the results of VQE and L-VQE with full precision energy computation in Tables \ref{tab:bestnonoiseSMO}--\ref{tab:averagenonoiseSMO}. In each iteration we evaluate the cost function exactly. In Table \ref{tab:bestnonoiseSMO} we report the best approximation ratio ($\rho_{\text{best}}$) achieved from the 10 runs using SMO for each graph. In Table \ref{tab:averagenonoiseSMO}, we report the average and standard deviation ($\rho_{\text{average}} \pm \sigma$) of the approximation ratio from the 10 runs for each graph.

\begin{table}
\centering
\caption{Best approximation ratio with full precision energy computation achieved by VQE and L-VQE using SMO. Comparing this table with Table \ref{tab:bestnoiseSMO}, L-VQE is clearly more robust to finite sampling errors.}\label{tab:bestnonoiseSMO}
\begin{tabular}{|c|c|c|c|c|c|}
\hline
\multirow{2}{*}{graph} & \multicolumn{2}{c|}{VQE $\rho_{\text{best}}$} & \multicolumn{3}{c|}{L-VQE $\rho_{\text{best}}$} \\
\cline{2-6}
      & 1 Layer & 2 Layer & 0 Layer & 1 Layer & 2 Layer \\
\hline
caveman & 1.0 & 1.0 & 0.95 & 1.0 & 1.0 \\\hline
caveman2 & 0.99 & 0.99 & 0.99 & 0.99 & 0.99 \\\hline
gaussian & 1.0 & 1.0 & 1.0 & 1.0 & 1.0 \\\hline
gaussian2 & 1.0 & 1.0 & 1.0 & 1.0 & 1.0 \\\hline
gnp1 & 1.0 & 0.94 & 1.0 & 1.0 & 1.0 \\\hline
gnp2 & 0.97 & 1.0 & 0.92 & 0.97 & 1.0 \\\hline
gnp3 & 1.0 & 1.0 & 1.0 & 1.0 & 0.99 \\\hline
gnp4 & 1.0 & 1.0 & 0.99 & 1.0 & 0.98 \\\hline
power & 0.94 & 0.93 & 0.98 & 0.99 & 1.0 \\\hline
power2 & 1.0 & 1.0 & 0.99 & 1.0 & 1.0 \\\hline
power3 & 0.96 & 0.96 & 0.9 & 1.0 & 1.0 \\\hline
random1 & 1.0 & 1.0 & 0.97 & 1.0 & 1.0 \\\hline
random2 & 1.0 & 1.0 & 0.98 & 1.0 & 1.0 \\\hline
random3 & 1.0 & 1.0 & 1.0 & 1.0 & 1.0 \\\hline
random4 & 0.98 & 0.97 & 1.0 & 1.0 & 1.0 \\\hline
windmill & 1.0 & 1.0 & 1.0 & 1.0 & 1.0 \\\hline
\end{tabular}
\end{table}

\begin{table*}
\centering
\caption{Average approximation ratio with full precision energy computation using SMO. Comparing this table with Table \ref{tab:averagenoiseSMO}, L-VQE is clearly more robust to finite sampling errors.}\label{tab:averagenonoiseSMO}
\begin{tabular}{|c|c|c|c|c|c|}
\hline
\multirow{2}{*}{graph} & \multicolumn{2}{c|}{VQE $\rho_{\text{average}} \pm \sigma$} & \multicolumn{3}{c|}{L-VQE $\rho_{\text{average}} \pm \sigma$} \\
\cline{2-6}
      & 1 Layer & 2 Layer & 0 Layer & 1 Layer & 2 Layer \\
\hline
caveman & 0.89 $\pm$ 0.12 & 0.92 $\pm$ 0.06 & 0.76 $\pm$ 0.11 & 0.9 $\pm$ 0.09 & 0.92 $\pm$ 0.09 \\\hline
caveman2 & 0.96 $\pm$ 0.04 & 0.96 $\pm$ 0.07 & 0.92 $\pm$ 0.06 & 0.99 $\pm$ 0.0 & 0.99 $\pm$ 0.0 \\\hline
gaussian & 0.94 $\pm$ 0.07 & 0.92 $\pm$ 0.08 & 0.81 $\pm$ 0.08 & 0.94 $\pm$ 0.09 & 0.85 $\pm$ 0.08 \\\hline
gaussian2 & 0.97 $\pm$ 0.05 & 0.99 $\pm$ 0.02 & 0.86 $\pm$ 0.1 & 0.99 $\pm$ 0.04 & 0.97 $\pm$ 0.04 \\\hline
gnp1 & 0.94 $\pm$ 0.04 & 0.89 $\pm$ 0.04 & 0.88 $\pm$ 0.08 & 0.92 $\pm$ 0.05 & 0.9 $\pm$ 0.05 \\\hline
gnp2 & 0.9 $\pm$ 0.05 & 0.94 $\pm$ 0.04 & 0.89 $\pm$ 0.03 & 0.92 $\pm$ 0.05 & 0.92 $\pm$ 0.05 \\\hline
gnp3 & 0.92 $\pm$ 0.08 & 0.88 $\pm$ 0.06 & 0.89 $\pm$ 0.06 & 0.9 $\pm$ 0.07 & 0.92 $\pm$ 0.06 \\\hline
gnp4 & 0.94 $\pm$ 0.03 & 0.95 $\pm$ 0.05 & 0.88 $\pm$ 0.07 & 0.96 $\pm$ 0.03 & 0.92 $\pm$ 0.06 \\\hline
power & 0.91 $\pm$ 0.04 & 0.9 $\pm$ 0.03 & 0.87 $\pm$ 0.07 & 0.92 $\pm$ 0.08 & 0.89 $\pm$ 0.08 \\\hline
power2 & 0.95 $\pm$ 0.04 & 0.93 $\pm$ 0.06 & 0.91 $\pm$ 0.08 & 0.94 $\pm$ 0.05 & 0.92 $\pm$ 0.08 \\\hline
power3 & 0.84 $\pm$ 0.08 & 0.89 $\pm$ 0.06 & 0.82 $\pm$ 0.08 & 0.91 $\pm$ 0.05 & 0.92 $\pm$ 0.05 \\\hline
random1 & 0.87 $\pm$ 0.14 & 0.97 $\pm$ 0.02 & 0.92 $\pm$ 0.06 & 0.97 $\pm$ 0.02 & 0.98 $\pm$ 0.02 \\\hline
random2 & 0.96 $\pm$ 0.04 & 0.95 $\pm$ 0.06 & 0.9 $\pm$ 0.06 & 0.96 $\pm$ 0.03 & 0.97 $\pm$ 0.02 \\\hline
random3 & 0.96 $\pm$ 0.08 & 0.95 $\pm$ 0.07 & 0.9 $\pm$ 0.11 & 0.97 $\pm$ 0.04 & 0.96 $\pm$ 0.04 \\\hline
random4 & 0.89 $\pm$ 0.06 & 0.91 $\pm$ 0.05 & 0.8 $\pm$ 0.12 & 0.94 $\pm$ 0.08 & 0.94 $\pm$ 0.06 \\\hline
windmill & 0.96 $\pm$ 0.06 & 0.94 $\pm$ 0.06 & 0.82 $\pm$ 0.12 & 0.96 $\pm$ 0.06 & 0.96 $\pm$ 0.06 \\\hline
\end{tabular}
\end{table*}

\subsubsection{Summary of VQE and L-VQE}\label{sec:experimentsvqesummary}

Across all instances we set the threshold of approximation ratio to 0.99, 0.95, and 0.90, respectively, and in Table \ref{tab:percentSMO} we report the percentage of the local optimizer runs that find the quantum state with a higher approximation ratio at the end of the algorithm. The rows in blue are the experiments with finite number of samples (i.e., the cost function is estimated by the mean of the samples), and the rows in white are the experiments with full precision energy computation (i.e., the cost function is evaluated exactly).

\begin{table}
\caption{Percentage of runs of local optimizers that reach a given approximation ratio: blue rows show results from experiments with the energy computed from a finite number of samples (mean of 2,000 samples); white rows are from experiments with full precision energy computation. The optimizer is SMO. With finite sampling errors, as the number of layers in the ansatz increases, results of VQE deteriorate. But for L-VQE, we achieve better results. Thus L-VQE is more robust to finite sampling errors compared with VQE.}\label{tab:percentSMO}
\begin{tabular}{|c|c|c|c|}
\hline

\rowcolor{lightgray}Approximation ratio $> 0.99$    & 0 Layer & 1 Layer & 2 Layer \\
\hline
\rowcolor{cyan}VQE           &    11.875\% & 0.625\% & 0.0\% \\
\hline
\rowcolor{cyan}L-VQE                   & 11.875\% & 29.375\% & 30.625\% \\
\hline
VQE         & 7.5\%    & 26.875\%   & 24.375\%     \\
\hline
L-VQE                  & 7.5\%    & 31.25\% & 27.5\%  \\
\hline

\rowcolor{lightgray}Approximation ratio $> 0.95$    & 0 Layer & 1 Layer & 2 Layer \\
\hline
\rowcolor{cyan}VQE            & 21.25\% & 33.75\% & 1.25\% \\
\hline
\rowcolor{cyan}L-VQE                     & 21.25\%    & 49.375\% & 48.125\% \\
\hline
VQE         & 18.75\% & 45.0\% & 48.125\% \\
\hline
L-VQE                     & 18.75\%   & 57.5\% & 58.125\%  \\
\hline

\rowcolor{lightgray}Approximation ratio $> 0.90$    & 0 Layer & 1 Layer & 2 Layer \\
\hline
\rowcolor{cyan}VQE     & 40.0\% & 55\% & 19.375\% \\
\hline
\rowcolor{cyan}L-VQE   & 40.0\% & 66.875\% & 67.5\%  \\
\hline
VQE         & 42.5\%    & 72.5\%   & 66.25\%     \\
\hline
L-VQE                    & 42.5\%    & 71.25\% & 71.25\%  \\
\hline
\end{tabular}
\end{table}

Intuitively, when we increase the size of the ansatz, the ansatz becomes more expressive, and we should have a better chance of finding the ground state. However, we can see that for VQE, when the cost function is estimated with finite number of samples, as the number of layers in the ansatz increase, the results deteriorate. But for L-VQE, as we increase the size of the ansatz, the results improve. Moreover, it is not practical to evaluate the energy exactly in applications when the size of the system gets larger. In L-VQE, by iteratively growing and reoptimizing the ansatz, we can achieve a higher probability of finding the ground state or a state that is sufficiently close to the ground state. By comparing the results of our L-VQE with or without full precision energy computation, we see no significant difference, which suggests that our approach is relatively robust to finite sampling errors. We expect that this behavior is caused by the complicated landscape of VQE spanned by many parameters. A large circuit, even though it is shallow, is hard to optimize if one does not use any mitigation strategies. L-VQE can be understood as one such strategy in which the circuit is carefully grown. This provides a good starting point for the optimization of a deeper circuit. Once the deeper circuit is initialized closer to the solution, the optimizer is less likely to hit the local minimum or spend more shots to escape from the neighborhood of the local minimum.

\begin{figure*}[!htbp]
	\centering
	\includegraphics[width=\linewidth]{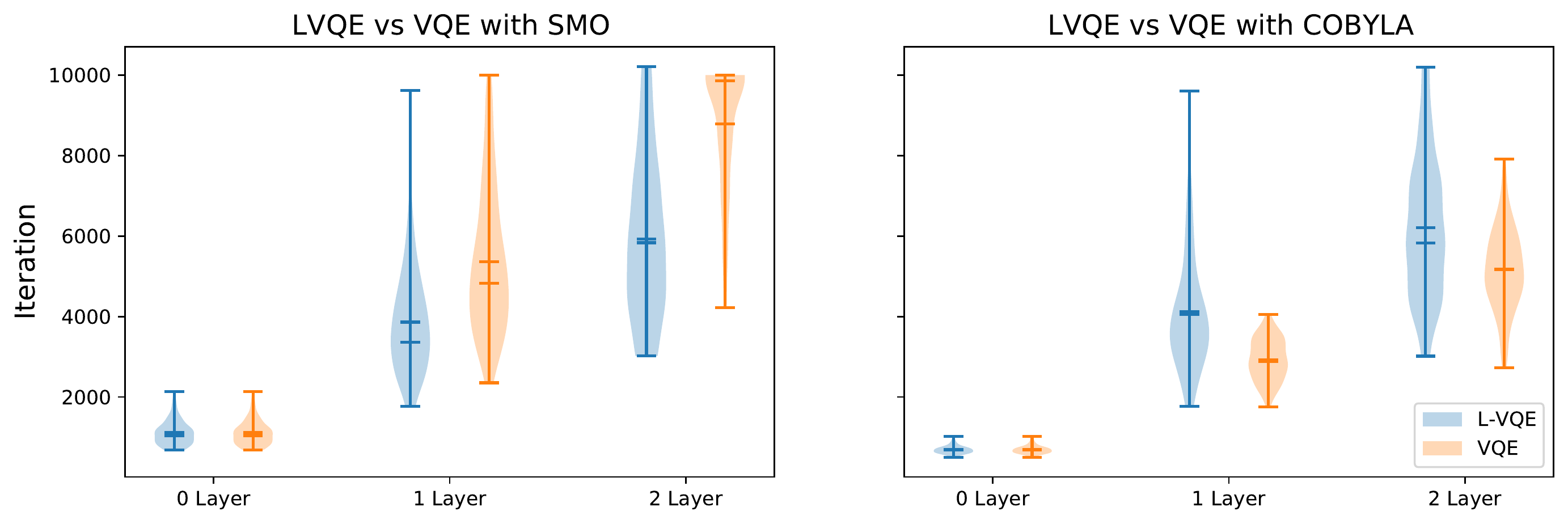}
	\caption{Average number of iterations needed for L-VQE and VQE on all 20 nodes graph (40 qubits), using SMO and COBYLA as optimizer. When using SMO as the optimizer, L-VQE needs fewer iterations in general. When using COBYLA, VQE needs fewer iterations. However, there is a tradeoff between the quality of the solution and the number of iterations.
	}
	\label{fig:lvqevsvqeiter}
\end{figure*}

\paragraph*{Additional evidence of L-VQE performance} We now provide additional evidence for the effects of reusing parameters and adding layers of the ansatz in L-VQE to complement the high-level statistics given in Tables \ref{tab:bestnoiseSMO} - \ref{tab:averagenonoiseSMO}. We observe that for the runs of experiments that start from the same initial Layer 0 ansatz, by reusing the parameters obtained from that ansatz, in most cases the results improve. Across all runs of the experiments, with finite samples, for 1 layer, 147 out of the 160 (91.88$\%$) runs find a state with a better or equal approximation ratio compared with the ansatz with 0 layer only. For 2 layers, 151 out of the 160 (94.38$\%$) runs find a state with a better or equal approximation ratio compared with the ansatz with 0 layers. Similarly, with full precision energy computation, across all runs of the experiments, for 1 layer, 150 out of the 160 (93.75$\%$) runs find a state with a better or equal approximation ratio compared with the ansatz with 0 layers. For 2 layers, 151 out of the 160 (94.38$\%$) runs find a quantum state with a better or equal approximation ratio compared with the ansatz with 0 layer. We present the violin plot of a representative instance \texttt{caveman} here in Fig. \ref{fig:lvqevsvqe}. Additional violin plots of the experiments for each graph can be found in Appendix \ref{app:vqeresult}. Finally, we present the average number of iterations needed for L-VQE and VQE on all 20 nodes graph (40 qubits) in Fig. \ref{fig:lvqevsvqeiter}. When using SMO as the optimizer, L-VQE needs fewer iterations in general. At the same time VQE requires fewer iterations if it used with COBYLA optimizer. It should be pointed out however that L-VQE obtains higher quality results. In near-term applications it is reasonable to use slightly more expensive technique if it gives more accurate results. As demonstrated here, this is the case with L-VQE vs VQE.

\begin{figure*}[!htbp]
	\centering
	\includegraphics[width=\linewidth]{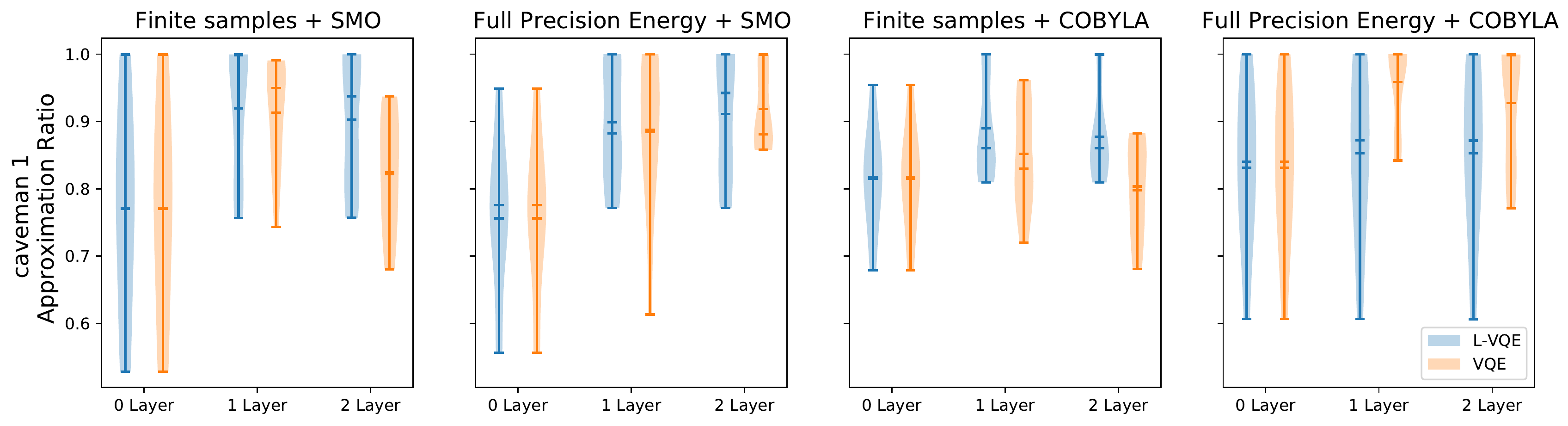}
	\caption{Violin plots of L-VQE vs VQE on graph \texttt{caveman}, finite samples vs full precision energy computation, using SMO and COBYLA as optimizer. The plots show the probability density of the results, with the kernel density estimator truncated to $(\min(\rho),\max(\rho))$ (since the approximation ratio cannot exceed 1). Comparing the results of L-VQE and with finite samples or full precision energy computation, we see no significant difference. But for VQE with finite samples, as the number of layers in the ansatz increases, the results deteriorate. L-VQE is relatively robust to sampling noise.
	}
	\label{fig:lvqevsvqe}
\end{figure*}

\subsection{Further Evidence of the Potential of L-VQE}\label{sec:experimentsevidence}

To provide further evidence of the potential of L-VQE, we present a scaling analysis of L-VQE in Section \ref{sec:experimentsevidencescale} and discuss the simulation results of L-VQE on a trapped ion noisy quantum simulator in Section \ref{sec:experimentsevidencenoisy}.

\subsubsection{Scaling analysis}\label{sec:experimentsevidencescale}
\begin{figure}[!htbp]
	\includegraphics[width=\linewidth]{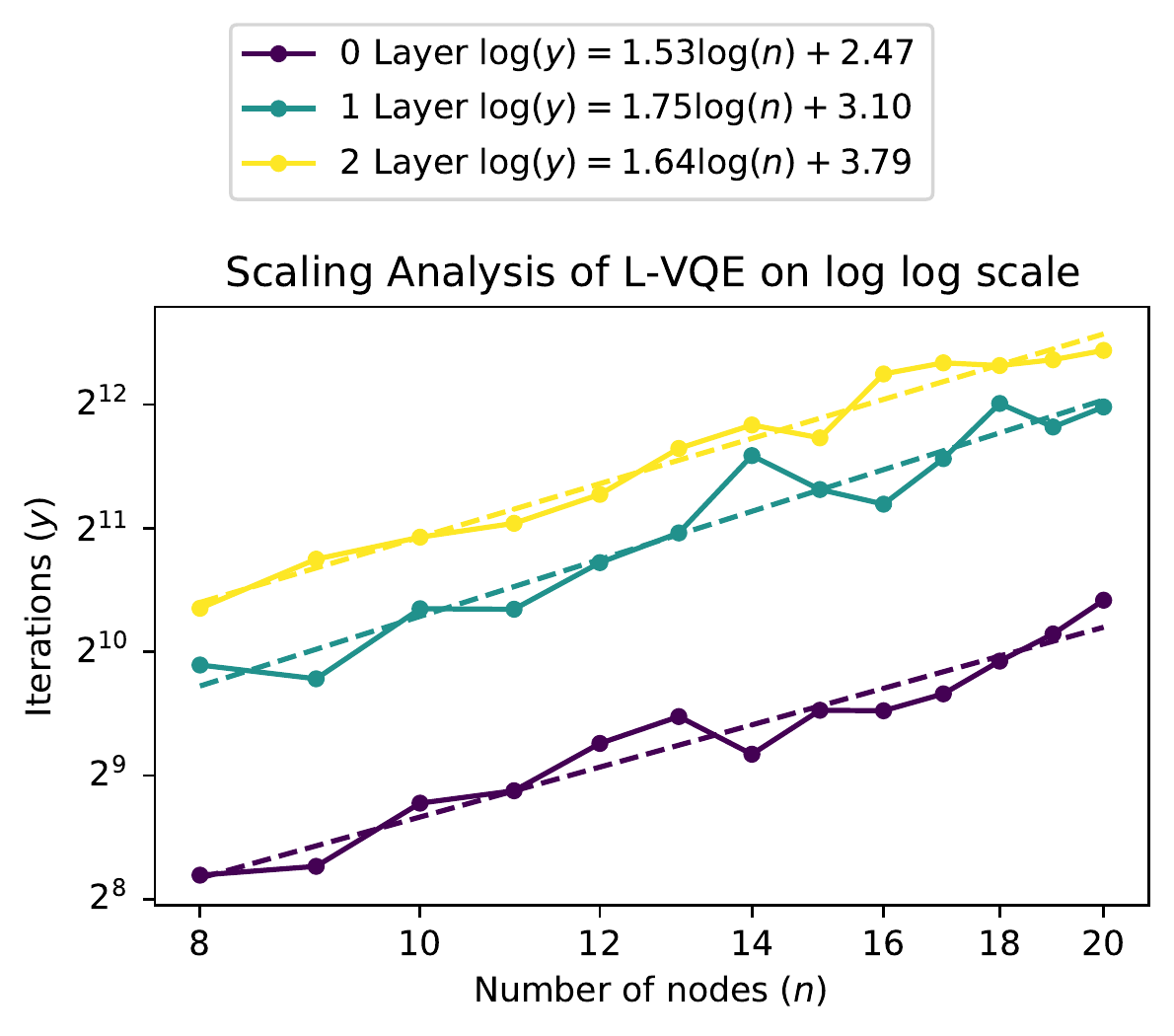}
	\caption{Average number of iterations until convergence scales up polynomially with respect to the size of the graph. }
	\label{fig:scalevqe}
\end{figure}

In this set of experiments, we generate random graphs with nodes ranging from 8 to 20. This means that in our application of finding a clustering up to 4 communities that maximize the modularity, we need to simulate qubits ranging from 16 to 40. For each graph and each approach, we run the experiments 10 times and record the average number of iterations needed for convergence of each graph. The results are shown in Fig. \ref{fig:scalevqe}. We can see that the number of iterations scales up polynomially as the number of nodes increases. Here, since within each iteration the number of $\rygate$ gates in the ansatz scales linearly with respect to the number of qubits needed (ansatz shown in Fig. \ref{fig:circuit1}), the number of parameters that need to be optimized therefore scales up linearly. In addition, the number of samples produced for evaluating the cost function is fixed as constant. Thus, the resources required for the entire algorithm scale polynomially. We point out, however, that our algorithm is heuristic by design and there is no guarantee of obtaining a solution with specified quality.

\subsubsection{Noisy simulations}\label{sec:experimentsevidencenoisy}

\begin{figure}[!htbp]
	\centering
	\includegraphics[width=\linewidth]{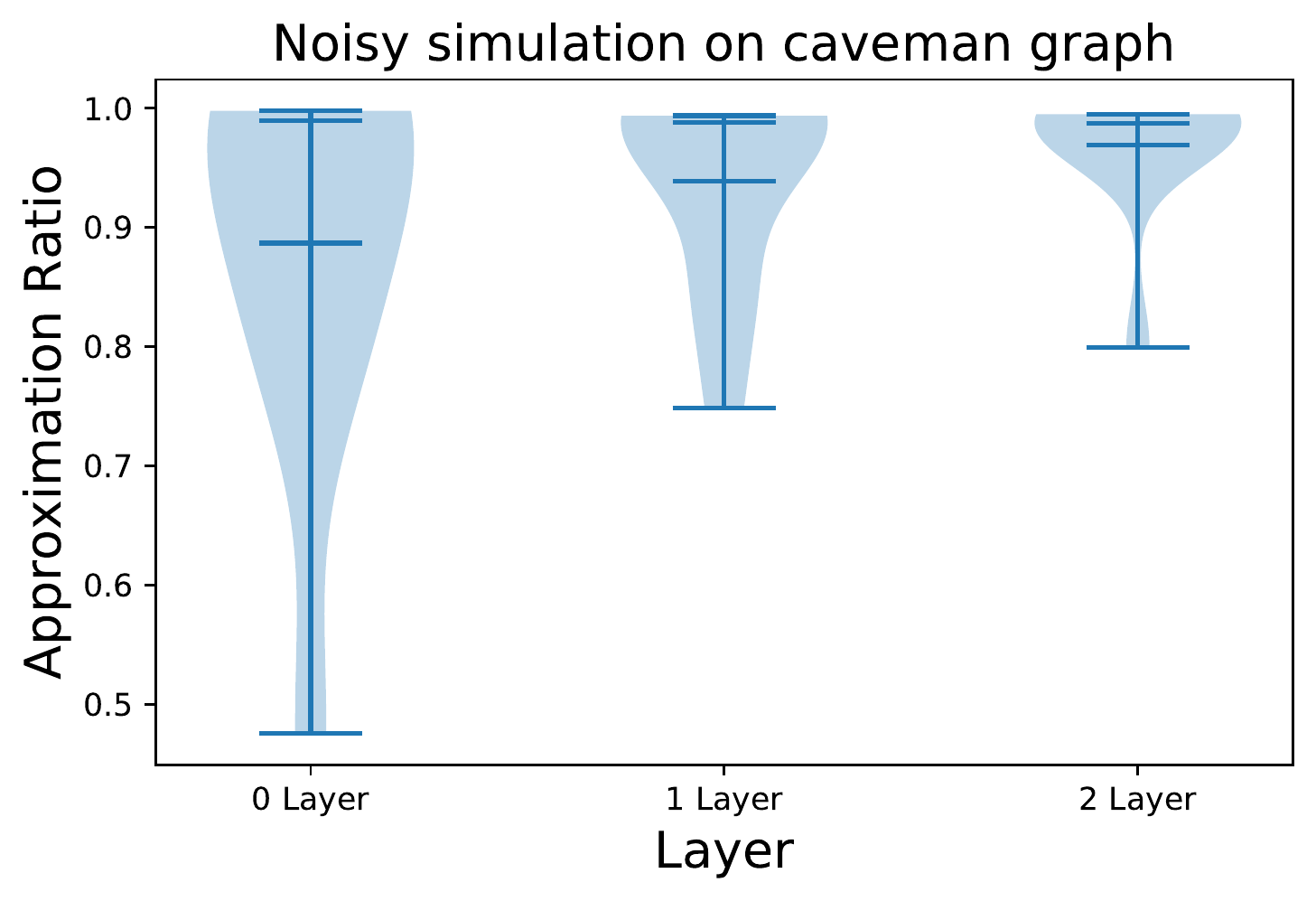}
	\caption{Violin plot of L-VQE performance on a trapped ion noisy quantum simulator. The plot shows the probability density of the results, with the kernel density estimator truncated to $(\min(\rho),\max(\rho))$ (since the approximation ratio cannot exceed 1). As the size of the ansatz increases, the probability of finding the ground state or a state that is sufficiently close increases.
	}
	\label{fig:distributionnoisy}
\end{figure}

The experiments described in the preceding sections are simulated in a setting that has no gate noise. For demonstration purposes, in the next set of experiments we also investigate the performance of L-VQE using a trapped ion noisy quantum simulator. We use realistic error rates in our simulations. Details of the noise model are given in Appendix\ref{app:noise_model}. We run the experiments on all 16 graphs, and show a representative instance \texttt{caveman} here, more results can be found in Appendix \ref{app:noisyresult} For L-VQE with Layer 1 and Layer 2, we run the experiments 10 times each. Fig. \ref{fig:distributionnoisy} gives a violin plot of the results. We observe that, in general, as the size of the ansatz increases, the probability of finding the ground state or a state that is sufficiently close increases. This suggests that L-VQE is also relatively robust to hardware noise and can be adapted to different quantum architectures.

\subsection{Entanglement vs no entanglement}\label{sec:experimentsentangle}
Our next experiment is aimed at understanding the role of entanglement in VQE. We use the same methodology as proposed in~\cite{nannicini2019performance}. That is, the experiment is based on replacing the entanglement gates $\cnotgate$ with a $\tgate$ gate acting on both qubits. Compared with previous work, with our simulator we can investigate the algorithm's performance on larger problems. We run the experiments on 4 graphs: (\texttt{caveman}, \texttt{gnp}, \texttt{random}, and \texttt{gaussian}). For each graph, we repeat the experiments 10 times with a different random seed. For the set of experiments with entanglement, we use the ansatz described in Fig. \ref{fig:circuit1} with Layer 0 and Layer 1. For the set of experiments without entanglement, we replace all $\cnotgate$ gates with a $\tgate$ gate acting on both qubits. The results are summarized in Table \ref{tab:entanglement}, where we report the percentage of runs that reach the approximation threshold 0.99, 0.95, and 0.90, respectively. As we can see from the results, under both cases, with and without full precision energy computation, using the ansatz with entanglement performs better than using the ansatz without entanglement.

\begin{table}
\centering
\caption{Percentage of experiments given the approximation ratio threshold}\label{tab:entanglement}
\begin{tabular}{|c|c|c|}
\hline
\rowcolor{lightgray} Approximation ratio $> 0.99$    & Entanglement & No entanglement\\
\hline
Finite samples        & 15\%    & 0\%     \\
\hline
Full precision energy computation       & 37.5\%    & 32.5\%   \\
\hline

\rowcolor{lightgray}Approximation ratio $> 0.95$    & Entanglement & No entanglement\\
\hline
Finite samples         & 45\% & 37.5\%   \\
\hline
Full precision energy computation  & 57.5\%    & 47.5\% \\
\hline
\rowcolor{lightgray}Approximation ratio $> 0.90$     & Entanglement & No entanglement \\
\hline
Finite samples & 65\%    & 60\%  \\
\hline
Full precision energy computation & 70\%    & 57.5\%     \\
\hline

\end{tabular}
\end{table}

\section{Conclusions and Discussion} \label{sec:conclusions}

Combinatorial optimization on near-term quantum devices is a leading candidate to demonstrate quantum advantage, and hybrid quantum-classical algorithms have been developed to solve this problem. In this work, we propose an iterative L-VQE approach inspired by VQE. We specifically studied the application of $k$-communities detection. In existing works, for a graph with $n$ nodes, solving the $k$-communities modularity maximization problem requires $kn$ qubits that encode the problem as an Ising model Hamiltonian. We propose a novel qubit-frugal formulation that requires only $n \left\lceil \log k \right\rceil$ qubits.

We compared the performance of L-VQE with QAOA, which is widely considered to be strong candidate for quantum advantage in applications with NISQ computers. However, the many-body terms in the Hamiltonian make it harder to implement in the QAOA setting. Moreover, the numerical results show that the optimization indeed gets harder, thus suggesting that L-VQE provides a practical alternative to QAOA for combinatorial optimization on noisy near-term quantum computers.

Unlike VQE, which has an ansatz fixed upfront, L-VQE starts from a simple and shallow hardware efficient ansatz with a small number of parameterized gates and then adds layers to the ansatz systematically. This strategy allows us to make the ansatz more expressive and reduces the optimization overhead. Our numerical results suggest that adding layers of the ansatz indeed increases the probability of finding the ground state or finding the state that is sufficiently close to the ground state. With finite samples, however, VQE is more likely to fail. We empirically observe L-VQE to be more robust to finite sampling errors, making it a promising approach for NISQ devices. We use matrix product state representation to perform large-scale simulations of the quantum circuits in MATLAB. Doing so allowed us to explore problems of larger size (simulations  up to 40 qubits and 352 parameters). We also studied the performance of L-VQE using a simulator of noisy trapped-ion quantum computer. The results suggest that our approach is relatively robust to hardware noise and can be adapted and generalized to different quantum architectures. Finally, we present numerical results of the role of entanglement in VQE. The results clearly show that the ansatz with entanglement performs better than the ansatz without entanglement.

Our results are the first indication that the introduction of additional entangling parameters in VQE for classical problems, as proposed in~\cite[Section V-B]{mcclean2020low}, break down the barriers in the optimization landscape, making it more convex and therefore more amenable to simple local outer-loop optimizers to find a minimum. This is in sharp contrast with the previous results of Nannicini~\cite{nannicini2019performance}, who did not observe any beneficial effects of entanglement. The difference in findings between our results and those presented in~\cite{nannicini2019performance} suggests the importance of the parameterization choice and the overall VQE procedure design to the success of such methods. We hope that this work will lead to even better algorithms to design ans\"atze for NISQ devices.

\section*{Acknowledgments}

We thank Jeffrey Larson for help with tuning APOSMM for QAOA parameter optimization. Clemson University is acknowledged for generous allotment of compute time on the Palmetto cluster. X.L., A.A., R.S., I.S. and Y.A. were supported in part with funding from the Defense Advanced Research Projects Agency (DARPA). R.S. and Y.A. were supported by Laboratory Directed Research and Development (LDRD) funding from Argonne National Laboratory, provided by the Director, Office of Science, of the U.S. Department of Energy under Contract No. DE-AC02-06CH11357. R.S. was supported by the U.S. Department of Energy, Office of Science, Office of Advanced Scientific Computing Research, Accelerated Research for Quantum Computing program. L.C. was supported by the Laboratory Directed Research and Development (LDRD) program of Los Alamos National Laboratory (LANL) under project number 20200056DR. LANL is operated by Triad National Security, LLC, for the National Nuclear Security Administration of U.S. Department of Energy (contract no. 89233218CNA000001). L.C. was also supported by the U.S. DOE, Office of Science, Office of Advanced Scientific Computing Research, under the Accelerated Research in Quantum Computing (ARQC) program.

\bibliographystyle{IEEEtran}
\bibliography{bib.bib}

\appendices

\section{QAOA experiments}
\label{app:qaoa}
\begin{figure}[!htbp]
\centering
  \begin{tikzpicture}[every node/.style={anchor=south west,inner sep=0pt}, x=1mm, y=1mm]
     \node (fig1) at (0,0)
      {\includegraphics[width=\linewidth]{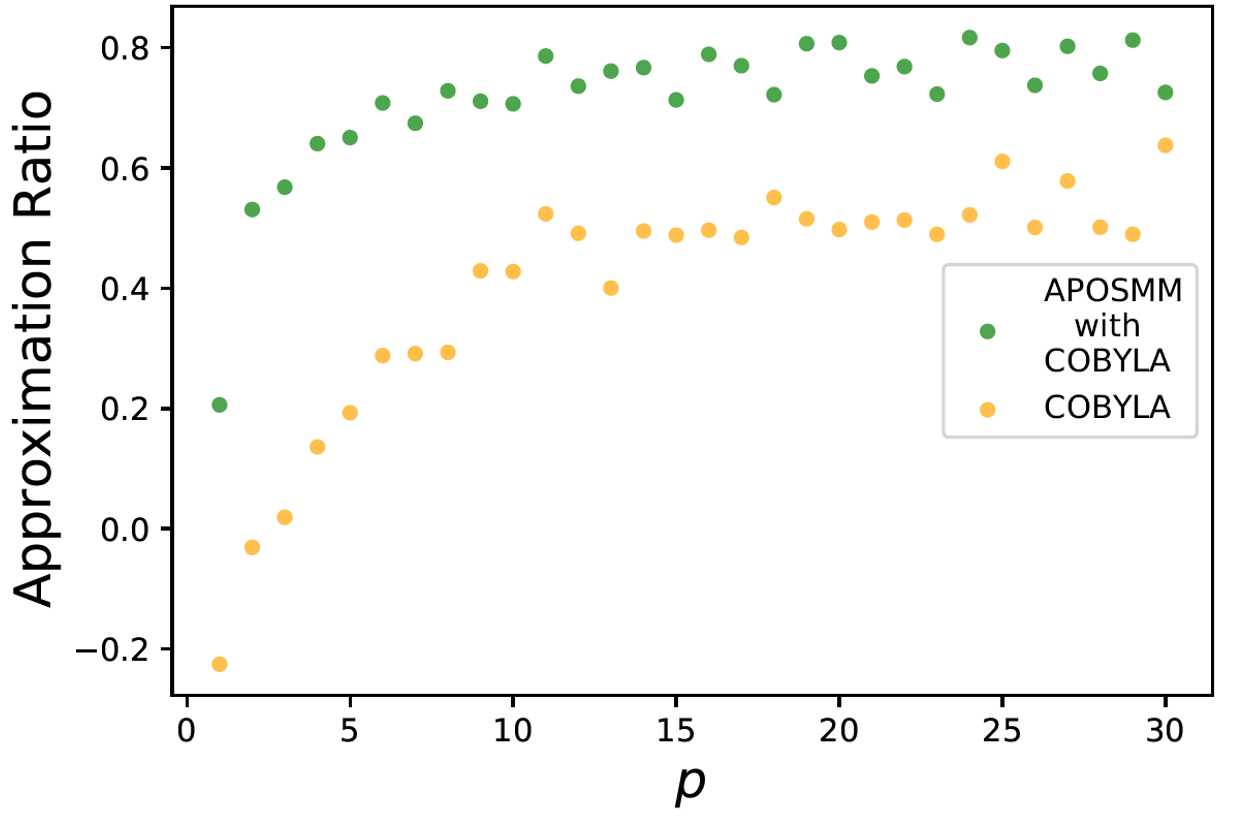}};
     \node (fig2) at (30, 10)
      {\includegraphics{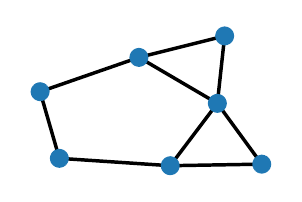}};
\end{tikzpicture}
\caption{Best approximation ratio QAOA found for the 7-node graph (shown in the inset) with $p$ ranging from 1 to 30. Even with the multistart method APOSSM to improve the optimizer COBYLA, we at most find an estimate of the ground state up to approximation ratio 0.817.}
\label{fig:qaoascale7_1}
\end{figure}

In this appendix we provide some additional details of the QAOA experiments on the \texttt{gnp random} graph with 7 nodes, shown in the inset of Fig. \ref{fig:qaoascale7_1}, simulating 14 qubits. The maximal modularity with up to 4 communities of this graph can be found by brute force (0.1790). We report the approximation ratio $\rho$ (defined in (\ref{eq:approximationratio})) found by QAOA in Fig. \ref{fig:qaoascale7_1}. We first run QAOA with $p$ ranging from 1 to 30 for 10 times for each $p$, using COBYLA to optimize the parameters. Each run is given a different random seed and is run until convergence. In Fig. \ref{fig:qaoascale7} we report the best approximation ratio we find from the 10 runs. Note that local optimizers such as COBYLA cannot guarantee to find the optimal parameters, especially as $p$ increases. This is the reason that the data points of approximation ratio do not grow monotonically with $p$. Therefore, to further improve the optimizer, we use the multistart method APOSMM with COBYLA, which uses a ensemble of local optimization solvers. We use COBYLA as the local optimization solver within APOSMM. We give APOSMM a limit of 30,000 iterations. The limit is chosen based on an empirical observation that with this parameter choice APOSMM will restart COBYLA for at least 10 times, usually much more. Using multistart method, the results improve compared with using only COBYLA. We observe that with this small graph, even if we increase $p$ up to 30, QAOA at most finds an estimate of the ground state up to approximation ratio 0.817.

\section{Trapped-ion quantum computer noise model} \label{app:noise_model}

In this Appendix we give details on the error model that was used in Section \ref{sec:experimentsevidencenoisy} of the main text. The error model is derived for near-term trapped-ion quantum computer with realistic error rates specified below. Ref.~\cite{trout2018simulating} develops error maps to model errors that accumulate during the execution of single qubit rotations $X(\theta) = e^{i\theta X}$, $Y(\theta) = e^{i\theta Y}$, $Z(\theta) = e^{i\theta Z}$ as well as during the execution of Molmer-S{\o}rens{\o}n gate $XX(\theta) = e^{i \theta X \otimes X}$. $X$, $Y$ and $Z$ denote spin-$\frac{1}{2}$ Pauli matrices. The quantum channels representing the noisy action of the above-mentioned gates take the following form:
\begin{eqnarray}
\mathcal{E}_X(\theta) &=&
\mathcal{D}(p_\mathrm{d}) \circ \mathcal{W}(p_\mathrm{dep}) \circ
\mathcal{R}_X(p_\alpha) \circ \mathcal{U}_X(\theta) \ ,\\
\mathcal{E}_Y(\theta) &=&
\mathcal{D}(p_\mathrm{d}) \circ \mathcal{W}(p_\mathrm{dep}) \circ
\mathcal{R}_Y(p_\alpha) \circ \mathcal{U}_Y(\theta) \ ,\\
\mathcal{E}_Z(\theta) &=&
\mathcal{D}(p_\mathrm{d}) \circ \mathcal{W}(p_\mathrm{dep}) \circ
\mathcal{R}_Z(p_\alpha) \circ \mathcal{U}_Z(\theta) \ ,\\
\mathcal{E}_{XX}(\theta) &=&
\big( \mathcal{D}(p_{\mathrm{d},1}) \otimes \mathcal{D}(p_{\mathrm{d},2}) \big) \circ \nonumber \\
 & & \big( \mathcal{W}(p_\mathrm{dep} ) \otimes \mathcal{W}(p_\mathrm{dep} ) \big) \circ \nonumber \\
 & & \mathcal{H}(p_\mathrm{xx}) \circ
 \mathcal{H}(p_\mathrm{h}) \circ
 \mathcal{U}_{XX}(\theta) \ .
\end{eqnarray}
Here, $\mathcal{U}_V (\theta)$ represents an ideal unitary evolution according to unitary $V$. That is,
\begin{equation}
\mathcal{U}_V (\theta) \rho = e^{-i\theta V}\rho e^{i\theta V} \ .
\end{equation}
$\mathcal{D}(p_d)$ is a dephasing channel defined as
\begin{equation}
\mathcal{D}(p_\mathrm{d}) \rho = (1 - p_\mathrm{d}) \rho
 + p_\mathrm{d} Z \rho Z \ .
\end{equation}
Note that in the definition of $\mathcal{E}_{XX}(\theta)$ we use separate dephasing channels for each qubit with (potentially different) error rates $p_{\mathrm{d},1}$, $p_{\mathrm{d},2}$.

Depolarizing channel $\mathcal{W}(p_\mathrm{dep})$ is defined as follows:
\begin{equation}
\mathcal{W}(p_\mathrm{dep}) \rho = (1 - p_\mathrm{dep}) \rho
 + \frac{p_\mathrm{dep}}{3} X\rho X
 + \frac{p_\mathrm{dep}}{3} Y\rho Y
 + \frac{p_\mathrm{dep}}{3} Z\rho Z \ .
\end{equation}

Imprecise rotation is implemented with $\mathcal{R}_V(p_\alpha)$. It is defined with
\begin{equation}
\mathcal{R}_V(p_\alpha) = (1-p_\alpha)\rho +
p_\alpha V^\dagger \rho V \ ,
\end{equation}
where $V = X,Y,Z$.

Finally, $\mathcal{H}(p_\mathrm{xx})$ represents the effects of two-qubit imprecise rotation and $\mathcal{H}(p_\mathrm{h})$ implements ion heating. The channel is defined in the following way:
\begin{equation}
\mathcal{H}(p) \rho =
(1-p) \rho + p (X\otimes X) \rho (X\otimes X) \ .
\end{equation}

We also model the effects of noisy initial state preparation. In our simulations, the perfect state $\rho_0 = \mathrm{diag}(1,0)$ is replaced with the state affected by depolarizing channel: $\mathcal{W}(p_\mathrm{dep})\rho_0$. Similarly, the measurement error is modeled with depolarizing channel. It is implemented by preceding the ideal POVM element with an action of depolarizing channel.

We used the following realistic values of noise rates:
\begin{eqnarray}
p_\mathrm{d} &=& 1.5 \times 10^{-4} \ , \nonumber \\
p_\mathrm{dep} &=& 8 \times 10^{-4} \ , \nonumber \\
p_{\mathrm{d},1} = p_{\mathrm{d},2} &=& 7.5 \times 10^{-4} \ , \nonumber \\
p_\alpha &=& 1 \times 10^{-4} \ , \nonumber \\
p_\mathrm{xx} &=& 1 \times 10^{-3} \ , \nonumber \\
p_\mathrm{h} &=& 1.25 \times 10^{-3} \ .
\end{eqnarray}

\section{Additional L-VQE and VQE simulation results}
\label{app:vqeresult}
\begin{figure*}[!htbp]
	\centering
	\includegraphics[width=\linewidth]{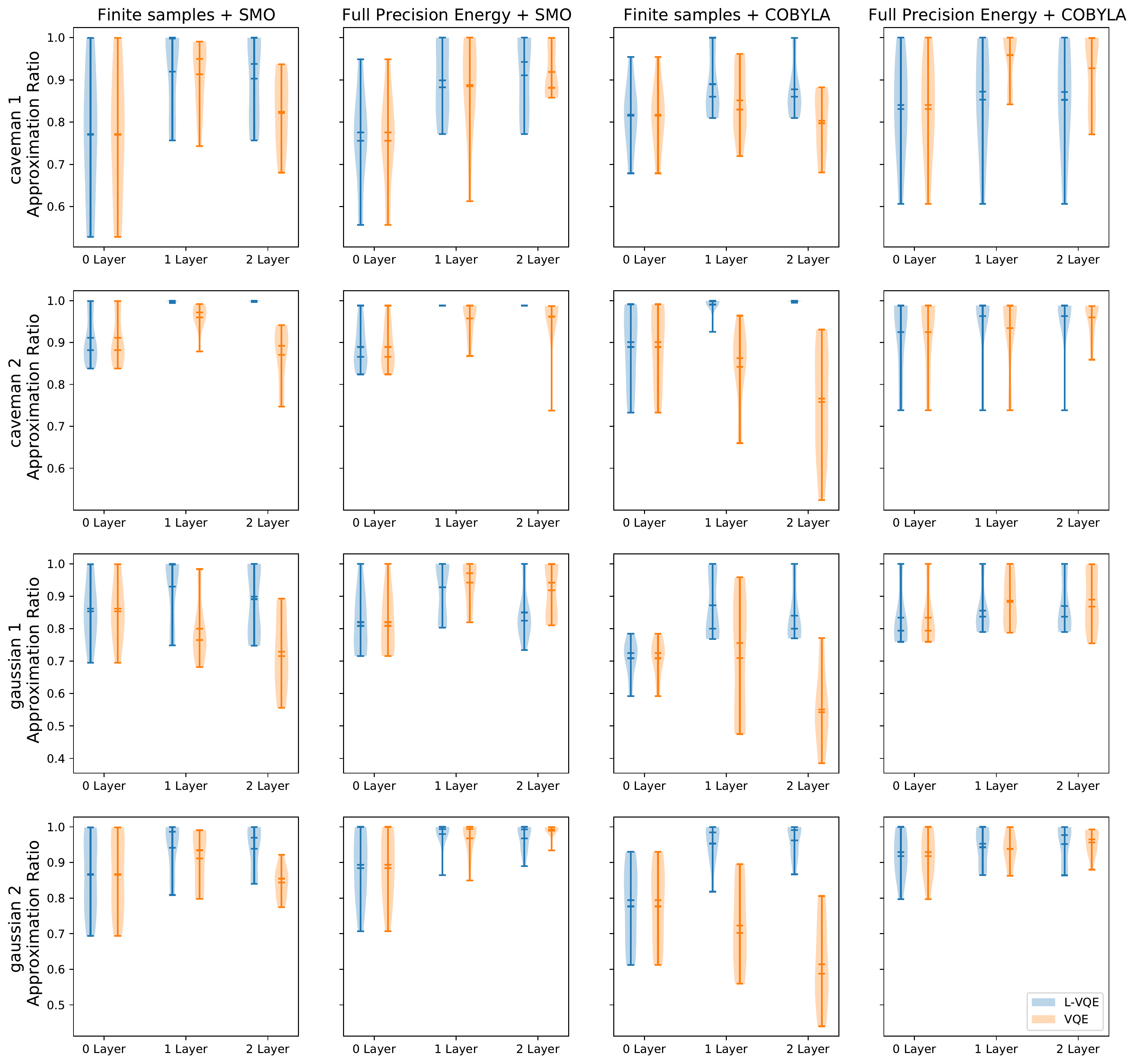}
	\caption{Violin plots of L-VQE vs VQE, finite samples vs full precision energy computation, using SMO and COBYLA as optimizer. The plots show the probability density of the results, with the kernel density estimator truncated to $(\min(\rho),\max(\rho))$ (since the approximation ratio cannot exceed 1). In general, as the number of layers in the ansatz increases, results of VQE deteriorate, but for L-VQE, we achieve better results.
	}
	\label{fig:lvqevsvqe1}
\end{figure*}

\begin{figure*}[!htbp]
	\centering
	\includegraphics[width=\linewidth]{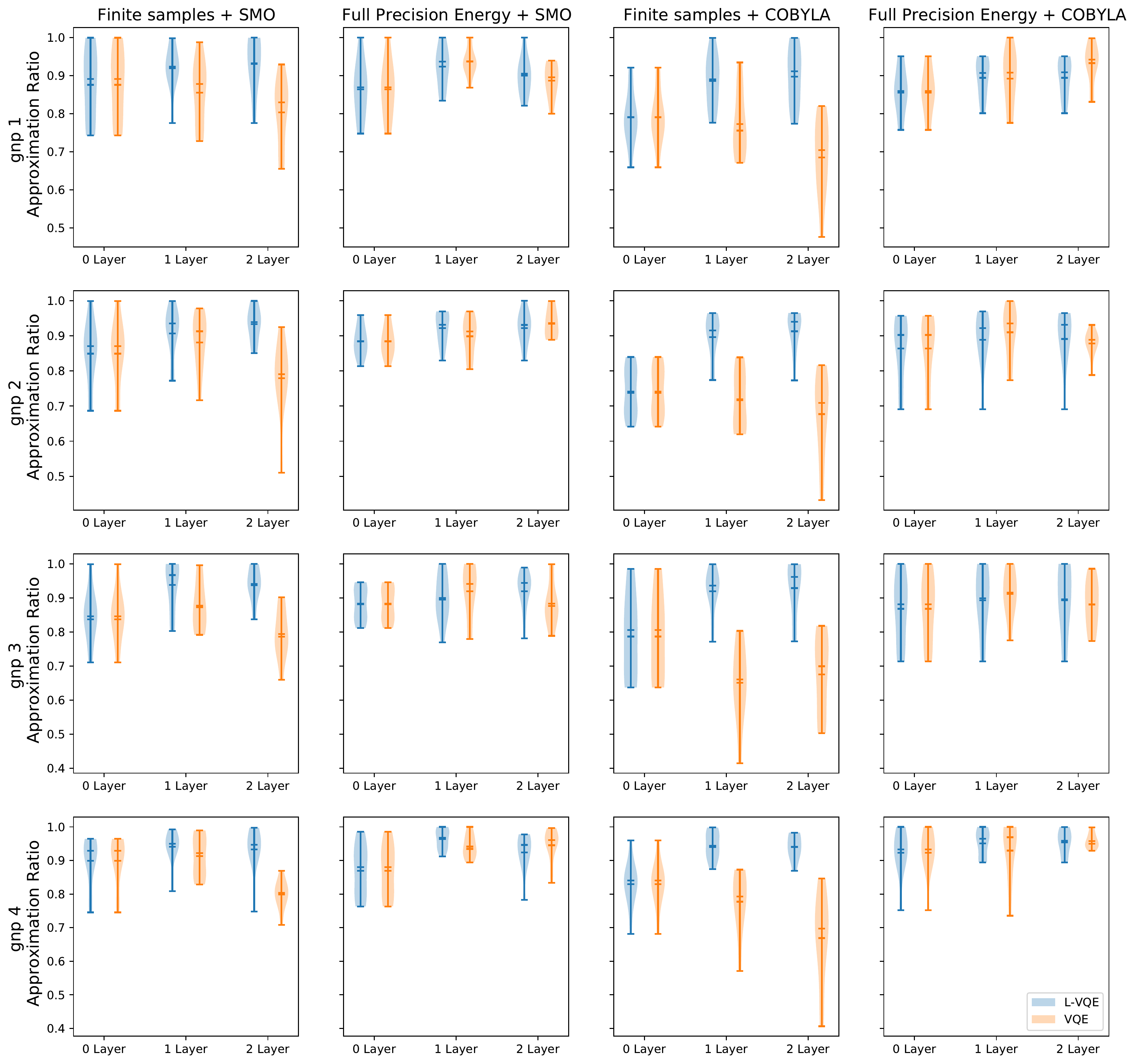}
	\caption{Violin plots of L-VQE vs VQE, finite samples vs full precision energy computation, using SMO and COBYLA as optimizer. The plots show the probability density of the results, with the kernel density estimator truncated to $(\min(\rho),\max(\rho))$ (since the approximation ratio cannot exceed 1). In general, as the number of layers in the ansatz increases, results of VQE deteriorate, but for L-VQE, we achieve better results.
	}
	\label{fig:lvqevsvqe2}
\end{figure*}

\begin{figure*}[!htbp]
	\centering
	\includegraphics[width=\linewidth]{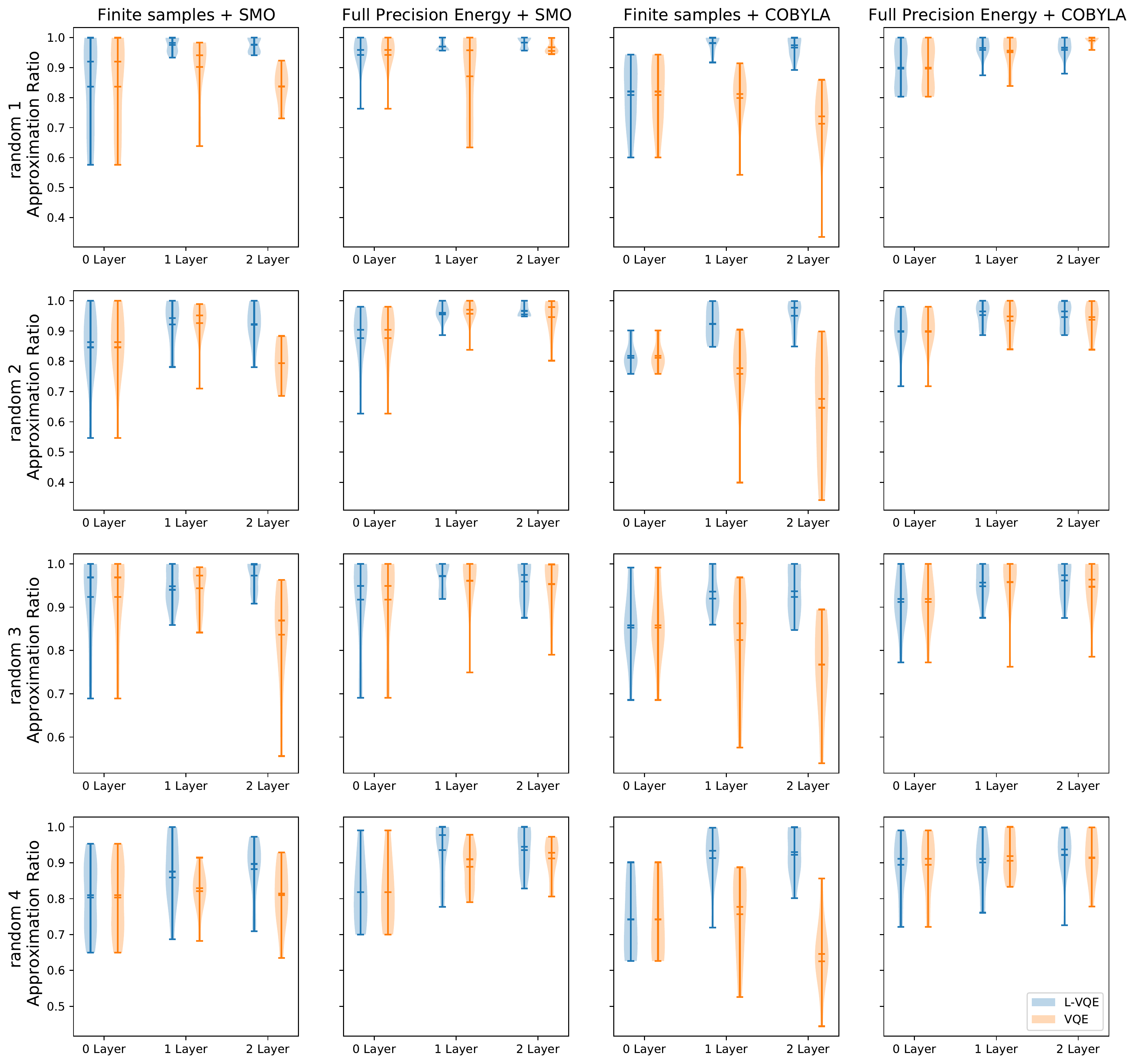}
	\caption{Violin plots of L-VQE vs VQE, finite samples vs full precision energy computation, using SMO and COBYLA as optimizer. The plots show the probability density of the results, with the kernel density estimator truncated to $(\min(\rho),\max(\rho))$ (since the approximation ratio cannot exceed 1). In general, as the number of layers in the ansatz increases, results of VQE deteriorate, but for L-VQE, we achieve better results.
	}
	\label{fig:lvqevsvqe3}
\end{figure*}

\begin{figure*}[!htbp]
	\centering
	\includegraphics[width=\linewidth]{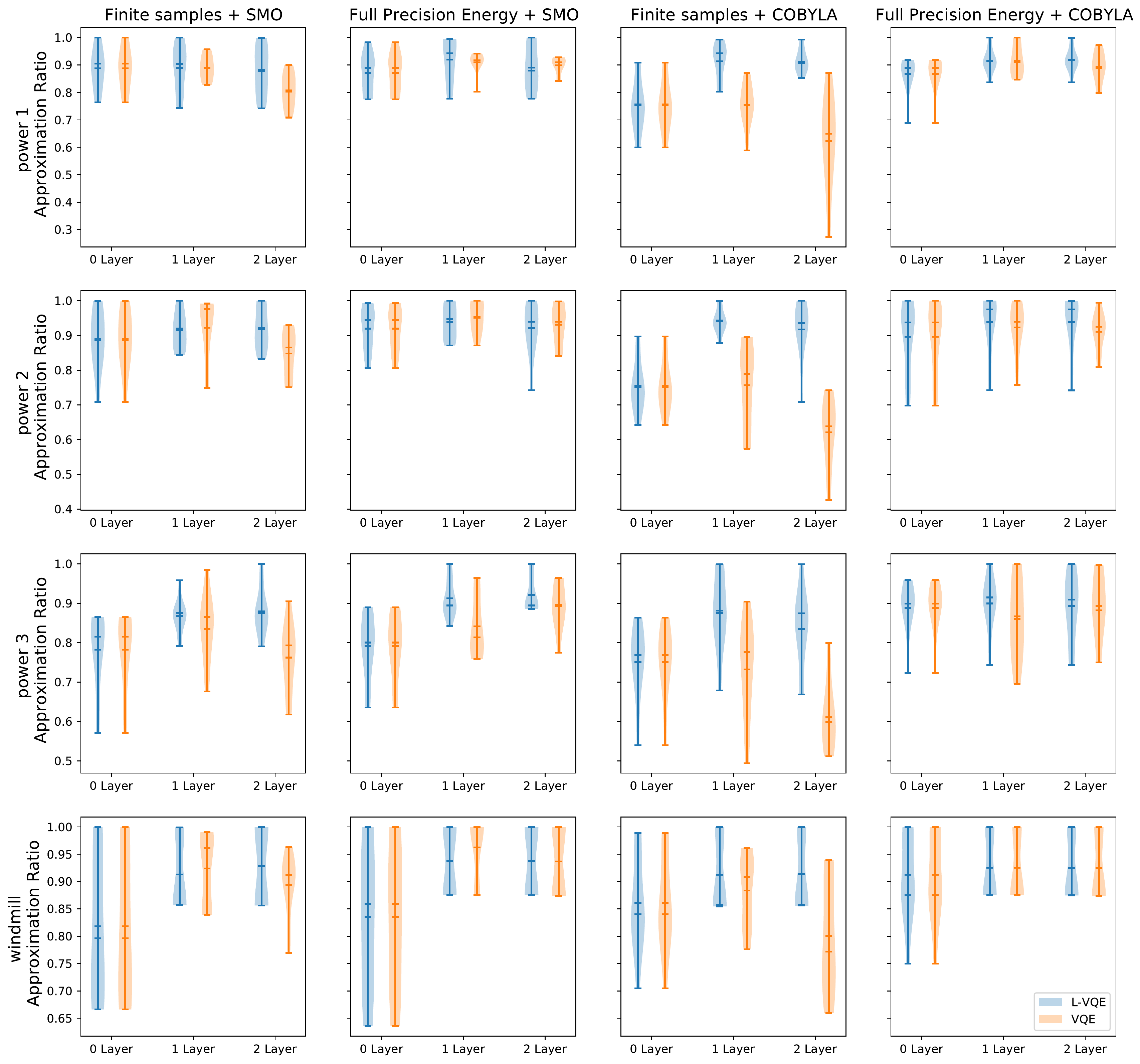}
	\caption{Violin plots of L-VQE vs VQE, finite samples vs full precision energy computation, using SMO and COBYLA as optimizer. The plots show the probability density of the results, with the kernel density estimator truncated to $(\min(\rho),\max(\rho))$ (since the approximation ratio cannot exceed 1). In general, as the number of layers in the ansatz increases, results of VQE deteriorate, but for L-VQE, we achieve better results.
	}
	\label{fig:lvqevsvqe4}
\end{figure*}

In this appendix we present more detailed simulation results of L-VQE and VQE using MPS simulator for all 16 graphs. Fig. \ref{fig:lvqevsvqe1}-\ref{fig:lvqevsvqe4} shows the violin plots of L-VQE vs VQE, finite samples vs full precision energy computation, using SMO and COBYLA as optimizer. In general, as the number of layers in the ansatz increases, results of VQE deteriorate, but for L-VQE, we achieve better results. This suggests that L-VQE is more robust to finite sampling errors compared with VQE.

\section{Additional L-VQE simulation results on trapped ion noisy simulator}
\label{app:noisyresult}

\begin{figure*}[!htbp]
	\centering
	\includegraphics[width=\linewidth]{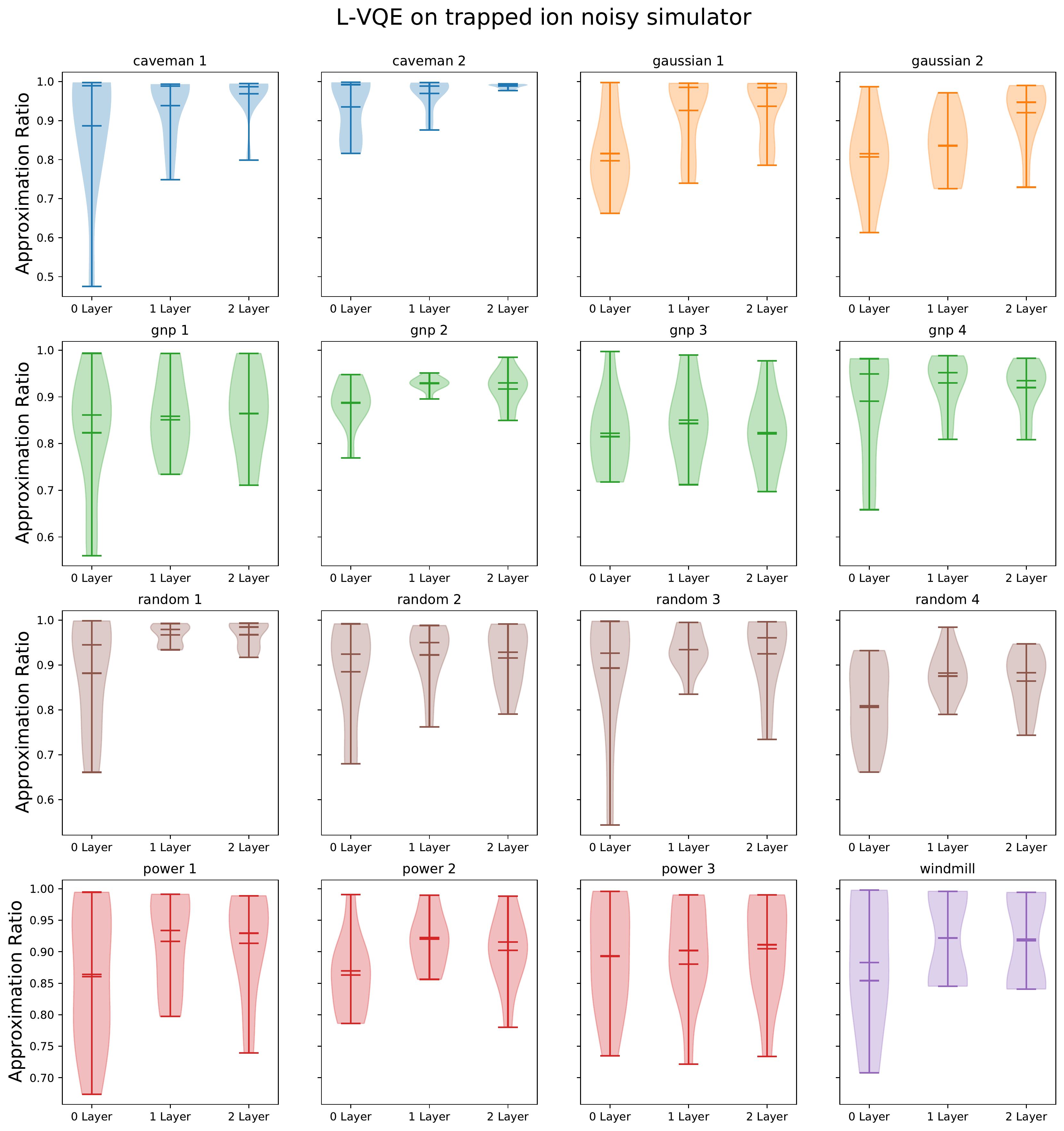}
	\caption{Violin plot of L-VQE performance on a trapped ion noisy quantum simulator. The plot shows the probability density of the results, with the kernel density estimator truncated to $(\min(\rho),\max(\rho))$ (since the approximation ratio cannot exceed 1). In general, as the size of the ansatz increases, the probability of finding the ground state or a state that is sufficiently close increases.
	}
	\label{fig:noisy}
\end{figure*}

In this appendix we present the simulation results of L-VQE using trapped ion noisy quantum simulator for all 16 graphs in Fig. \ref{fig:noisy}. In general, as the size of the ansatz increases, the probability of finding the ground state or a state that is sufficiently close increases. Therefore suggests that L-VQE is relatively robust to hardware noise.

\small
\vfill
\framebox{\parbox{\linewidth}{The submitted manuscript has been created by
UChicago Argonne, LLC, Operator of Argonne National Laboratory (``Argonne'').
Argonne, a U.S.\ Department of Energy Office of Science laboratory, is operated
under Contract No.\ DE-AC02-06CH11357.  The U.S.\ Government retains for itself,
and others acting on its behalf, a paid-up nonexclusive, irrevocable worldwide
license in said article to reproduce, prepare derivative works, distribute
copies to the public, and perform publicly and display publicly, by or on
behalf of the Government.  The Department of Energy will provide public access
to these results of federally sponsored research in accordance with the DOE
Public Access Plan \url{http://energy.gov/downloads/doe-public-access-plan}.}}

\EOD

\end{document}